\documentclass[traditabstract]{aa}  

\usepackage{graphicx}
\usepackage{txfonts}
\usepackage{xcolor}
\usepackage{float}
\usepackage{multirow}
\usepackage{booktabs}
\usepackage{soul}
\usepackage{ulem}
\usepackage[switch]{lineno}
\begin{document}

\title{Asymptotic giant branch stars in the eROSITA-DE eRASS1 catalog}

   \author{M.~A.\ Guerrero
          \inst{1}
          \and
          R.~Montez Jr.\ 
          \inst{2}
          \and
          R.~Ortiz
          \inst{3}
          \and
          J.~A.~Toal\'{a}\inst{4}
          \and
          J.~H.\ Kastner\inst{5,6}
          }

   \institute{
   Instituto de Astrof\'{i}sica de Andaluc\'{i}a, CSIC, Glorieta de la Astronom\'{i}a S/N, Granada, E-18008, Spain \\
         \email{mar@iaa.es}
         \and
    Smithsonian Astrophysical Observatory, Cambridge, MA 02138, USA 
    \and
    Escola de Artes, Ci\^encias e Humanidades, USP, Av. Arlindo Bettio 1000,
03828-000 S\~ao Paulo, Brazil 
    \and
    Instituto de Radioastronom\'{i}a y Astrof\'{i}sica, Universidad Nacional Aut\'{o}noma de M\'{e}xico, 58089 Morelia, Michoac\'{a}n, Mexico
    \and
    School of Physics and Astronomy and Laboratory for Multiwavelength Astrophysics, Rochester Institute of Technology, Rochester, NY, USA
    \and
    Chester F.\ Carlson Center for Imaging Science, Rochester Institute of Technology, USA
    }

\date{February 2024}

\abstract{
Asymptotic giant branch (AGB) stars are not expected to be X-ray-emitters, yet a small fraction of them, the so-called X-AGBs, show X-ray emission that can be attributed to
coronal activity of a companion or accretion onto one. 
}{
By searching the recently released SRG eROSITA-DE eRASS1 source catalog, we aim to increase the sample of known X-AGBs and investigate their X-ray and far-UV properties. 
So far, 36 X-AGBs have been reported, which includes 21 previous detections from ROSAT RASS, Chandra, and XMM-Newton and 15 recent detections from eROSITA eRASS1.
}{
We cross-correlated the position of sources in the eROSITA-DE eRASS1 catalog with the largest available samples of AGB stars in order to find possible X-ray counterparts. 
We carefully checked the possible counterparts by comparing X-ray end near-IR $K$ images, disregarding those affected by optical loading, those found to be diffuse sources, or those simply considered unreliable positional associations. 
}{
We have found seven high-confidence X-AGBs and another seven possible ones. 
Accounting for previous X-ray detections, the sample of X-AGBs is increased by 11 new sources, increasing the sample of X-AGBs from 36 up to 47.  
Adding these sources to previous eROSITA-DE eRASS1 X-AGB detections, eROSITA has so far discovered 26 new X-AGBs, more than doubling the number of known X-AGBs.
This demonstrates eROSITA's capability to detect X-AGBs despite the challenge posed by the optical loading caused by their near-IR brightness, which makes the X-ray detection untrustworthy in a number of cases.
}{
The eRASS1 X-AGBs tend to have a higher X-ray luminosity than that of previously detected X-AGBs, suggesting a bias toward brighter sources that is very likely due to the short exposure time of eRASS1 sources. 
A comparison of the X-ray and far-UV luminosity of X-AGBs with those of X-ray-emitter symbiotic stars (X-SySts) revealed an overlap in the X-ray luminosity range $10^{29.5} < L_\mathrm{X}$ (erg s$^{-1}$) $< 10^{33.0}$.
The average higher X-ray luminosity of X-SySts AGBs ($\approx 10^{32}$ erg~s$^{-1}$) can be interpreted as X-ray emission arising from a boundary layer between an accretion disk and a white dwarf, whereas the average lower X-ray luminosity ($\approx 5\times10^{30}$ erg~s$^{-1}$) of X-AGBs would arise from an accretion disk around main-sequence or subgiant F-K companion stars.
} 

\maketitle
%

\section{Introduction}

Asymptotic giant branch (AGB) stars are bright cool giant stars with an inert core of carbon and oxygen, an inner shell where helium is burned into carbon, an outer shell where hydrogen is burned into helium, and an extended hydrogen-rich envelope.  
AGB stars are the immediate progenitors of planetary nebulae (PNe), and as such, they hold the keys to understanding their formation.  
Most PNe are believed to be shaped by binary interactions \citep{DeMarco2009}, and thus AGB stars have become a target for companion searches.

The discovery of a companion to an AGB star, however, is only straightforward when it is a member of a symbiotic star (SySt), a close binary system composed by a red giant and a white dwarf (WD). 
In these systems the components are so close to each other that mass transfer takes place from the giant primary to the secondary WD, very likely through an accretion disk around the latter \citep{Merc+2024}.
Thus, although the secondary cannot generally be visually detected, the spectrum of a SySt usually contains \citep[in addition to absorption lines produced at the atmosphere of the cooler red giant; e.g.,][]{Schmidt+2006} emission lines formed in the accretion disk that unveils the presence of the companion.  
On the other hand, if a binary system consists of an AGB star and a low- or intermediate-mass main-sequence or subgiant companion, the detection of the secondary is cumbersome because (1) the high luminosity of the AGB star overshines its companion, (2) no emission lines from an accretion disk are detected, and (3) the radial pulsations of the AGB star hamper the detection of the radial velocity variations caused by the mutual orbital motion.

Alternatively, the binarity of an AGB star can be revealed at high energies, particularly in X-rays.  
Single AGB stars are expected to be X-ray quiet because they are slow rotators and thus unlikely to have strong surface magnetic fields to support a corona \citep{Ayres+1981,LH1979}. 
Therefore, AGB stars with X-ray counterparts (hereafter X-AGBs) are believed to be members of binary systems, particularly those with X-ray luminosity in excess of a few times $10^{29}$ erg~s$^{-1}$ \citep{SK2003}, that is, most X-AGBs \citep{Sahai2015,OG2021}. 
The X-ray emission would originate either in the corona of their companion or in an accretion disk around a secondary, such as those around the degenerate WD component of X-ray-emitting SySts (hereafter X-SySts). 
Such an example is Y\,Gem, an X-AGB showing far-UV excess and a double-peaked X-ray spectrum \citep{OG2021} that has recently been proposed to actually be a SySt \citep{Yu+2022}.

To date, a growing number of AGB stars with X-ray counterparts have been reported in the literature \citep{Jorissen1996,Hunsch1998,KS2004,Ramstedt2012,Sahai2015}, making X-AGBs an interesting class of objects in the zoo of X-ray sources. 
A recent search for X-AGBs \citep{OG2021} carried out a cross-correlation between the 4XMM-DR9 catalog and various lists of AGB stars (exhibiting high mass-loss rates; nearby, listed in the {\it HIPPARCOS} catalog; showing various [C/O] relative abundances)
and found eight new objects, increasing the number of known X-AGBs at that time to 26, although five of them were later found to not be AGB stars \citep[e.g., DT\,Psc and HD\,35155 in][now known to be SySts]{Jorissen1996}. 
Their X-ray spectra can be described by absorbed optically thin plasma emission models with a wide interval of plasma temperatures, from 4 MK to 117 MK, which is similar to that displayed by X-SySts \citep[see][and references therein]{Merc2019}. 
On the other hand, the luminosity of X-AGBs is typically in the range $10^{-8} < L_\mathrm{X}/L_{\rm bol} < 10^{-6}$ ($10^{29} \sim 10^{31} L_{\odot}$), whereas X-SySts generally exhibit higher values, $10^{-7} < L_\mathrm{X}/L_{\rm bol} < 10^{-2}$ ($10^{30} \sim 10^{33} L_{\odot}$). 
The X-ray emission from X-AGBs was interpreted by \citet{OG2021} as being a consequence of binarity, especially because of the positive correlation between X-ray and far-UV luminosity. 


The aim of the extended ROentgen Survey with an Imaging Telescope Array (eROSITA) on board the Russian-German Spectrum-Roentgen-Gamma (SRG) observatory is to survey the whole sky at a depth much greater than the available ROSAT All-Sky Survey (RASS).  
The recent release of the first eROSITA German catalog, the so-called eROSITA-DE eRASS1 (hereafter eRASS1), provides a unique opportunity to increase the sample of X-AGBs.  
We present here a search of the eRASS1 catalog for point-source X-ray counterparts of AGB stars. 
After the first release of the eROSITA catalog, \citet{Schmitt2024} identified 15 new X-ray sources associated with very red giant stars,\footnote{\citet{Schmitt2024} considered R\,Hor, U\,Men, and R\,LMi to be spurious detections caused by optical loading of the X-ray detectors.} namely DR\,Eri, BH\,Eri, UU\,Ret, CD$-$51\,1503, CPD\,84$-$86, BD$+$32\,1528, TYC9506-1836-1, MV\,Hya, EH\,Leo, HD\,84048, WW\,Crt, TW\,Cen, IRAS\,19204$-$3959, HD\,181817, and HD\,199203, as well as the additional detection of the previously known X-AGB CI\,Hyi \citep{Sahai2015} among a sample of {\it Gaia} DR3 sources closer than 1300 pc.  
These X-AGBs show $L_\mathrm{X}$ in the range $\simeq 2\times10^{30}$ to $\simeq 2\times10^{31}$ erg~s$^{-1}$. 
Their X-ray emission is attributed to unseen companions of these AGB stars, even though just a few of them have been confirmed to be binaries by other means.
The total number of X-AGBs identified so far amounts to 36. 

In this paper, we search the eRASS1 catalog for X-ray counterparts of AGB stars among a larger sample of AGB stars.  
This paper is organized as follows: 
In Section~\ref{sec:obs}, we describe the sample of AGB stars, the eRASS1 catalog, and our search procedure. 
The list of potential X-AGBs is presented and discussed in Section~\ref{sec:analysis}. 
The discussion of our results is presented in Section~\ref{sec:discussion}. 
Finally, our conclusions are presented in Section~\ref{sec:conclusions}.

\section{The sample of asymptotic giant branch stars and the eROSITA eRASS1 catalog}
\label{sec:obs}

\subsection{The sample of asymptotic giant branch stars}
\label{subsec:sample}

Asymptotic giant branch stars are recognized by their high luminosity, late spectral type, strong emission in the infrared, and the enrichment of their atmosphere by elements generated during the third dredge-up episode, such as carbon and 
$s$-process elements.
A significant fraction of AGBs cannot be detected at visual wavelengths and are only detectable at infrared wavelengths because their optically thick circumstellar dust shells (CDSs) block the visual radiation emitted by
their photosphere.

Most samples of AGB stars in the literature are based on the properties described above. For example, the vast majority of the 2341 stars in the catalog of OH/IR stars of \citet{EB2015} are in the AGB phase, with a minor fraction of supergiants. 
In the cases where the abundance ratio [C/O] exceeds unity, other molecular species can be formed in the CDS besides OH and SiO, such as HCN, for example. The catalog of CO and HCN observations by \citet{Loup1993} contains 184 O-rich, 205 C-rich, and 9 S-type AGB stars (the latter showing [C/O] $\approx 1$). More recently, \citet{SH2017} compiled a new  list of AGB stars based on $K$+[IRAS] color-color diagrams, which was eventually updated by \citet{Suh2021} using IR photometry obtained by various surveys (e.g., near-IR 2MASS, mid-IR {\it MSX} and {\it WISE}, and far-IR {\it IRAS} and {\it AKARI}). The latest version of their catalog also includes 6528 large amplitude, regularly pulsating AGB stars (i.e., ``Mira-type'' variables) located in the Galactic bulge detected during the OGLE project \citep{Soszynski2013}.
In spite of all attempts to compile a definitive, updated list of Galactic AGB stars, there are still a significant number of objects that have been overlooked. For example, the following AGB stars with X-ray emission \citep{Sahai2015,OG2021} are not listed in the catalog by \citet{Suh2021}: CI\,Hyi, CD$-$33$^{\circ}$2309, CD$-$38$^{\circ}$3905, CSS\,1244, DH\,Eri, and EY\,Hya, among others.

In the present work, we have considered four samples of AGB stars: (1) the full New Catalog of Asymptotic Giant Branch Stars in Our Galaxy \citep{Suh2021}, amounting to 11,209 O-rich and 7,172 C-rich stars; 
(2) the sample of 469 AGB stars examined by Montez et al.\ (2017);
(3) the 13 previously detected X-AGBs by \citet{OG2021} and \citet{Sahai2015}; and (4) 10,820 Miras, SR- and Lb-type variables listed in the General Catalog of Variable Stars \citep[{\it GCVS},][]{Samus2017} showing any of the following characteristics, a Mira or semi-regular variable with P$>100^{\rm d}$, a spectral type later than M3{\sc iii}, and being a C- or S-type.
We reckon, as shown later in Sect.~\ref{subsec:cross} during the identification of eRASS1 X-ray counterparts of AGB stars, that the samples of AGB stars used here are not completely pure, 
meaning that they include non-AGB contaminant sources such as SySts and red dwarfs, among others.

\subsection{The eROSITA eRASS1 catalog}

Launched on 2019 July 13, eROSITA is a Russian-German mission that consists of an array of seven X-ray telescopes. 
It offers wide field and high-throughput X-ray spectroscopic imaging in the 0.2-8 keV energy range with an energy resolution of $\simeq$80 eV at 1.5 keV and an imaging resolution with a half energy width (HEW) $\simeq26^{\prime\prime}$ \citep{Predehl+2021}.

The initial 184-day all-sky survey of eROSITA was performed from 2019 December 12 to 2020 June 11.  
The data rights are split by Galactic longitude ($l$), with the ``eastern'' ($l < 180^\circ$) data for the Russian consortium and the ``western'' ($l > 180^\circ$) data for the German consortium.
The first data release of the western Galactic hemisphere of the eROSITA all-sky survey (the eROSITA-DE eRASS1; hereafter eRASS1) was made public on 2024 January 31 \citep[][]{Merloni+2024}.  
It has a typical average flux sensitivity 
limit of $5\times10^{-14} {\rm ~cm}^{-2} {\rm ~s}^{-1}$ that improves at high ecliptic latitudes due to the all-sky survey scanning pattern and an average spatial resolution with HEW $\simeq30^{\prime\prime}$. 
The eRASS1 data provides a suite of calibrated data products and source catalogs produced with the eROSITA standard data processing pipeline. 
In this work, we used the eRASS1 Main Catalog, which includes calibrated event lists and count numbers, count rates, flux estimates, detection likelihood, observing time, and quality flags for 903521 point sources and 26682 extended sources detected in the 0.2-2.3 keV band. 
The reader is referred to \citet{Predehl+2021} and \citet{Merloni+2024} for complete information on eROSITA and eRASS1, respectively.

It is relevant to note here that the eROSITA CCD detectors are prone to optical loading caused by the accumulation of low-energy photons within a CCD pixel over the frame integration time of 50 ms for bright optical/UV sources. 
This results in false, predominantly soft X-ray events.  
The adopted brightness limits of eRASS1 are $B$, $V$, $G \leq$ 4.5 mag and $J \leq$ 3 mag for tagging a source as potentially being the result of optical loading.  
If one or more of these criteria are fulfilled, the \texttt{FLAG\_OPT} column entry of the eRASS1 catalog is set to one. 
It is still worthwhile to remark that AGB stars can not only be bright, particularly at IR wavelengths, but also variable on timescales of months to years, adding special concern to spurious X-ray detections caused by optical load.

\begin{figure}
\centering
\includegraphics[width=1.0\linewidth]{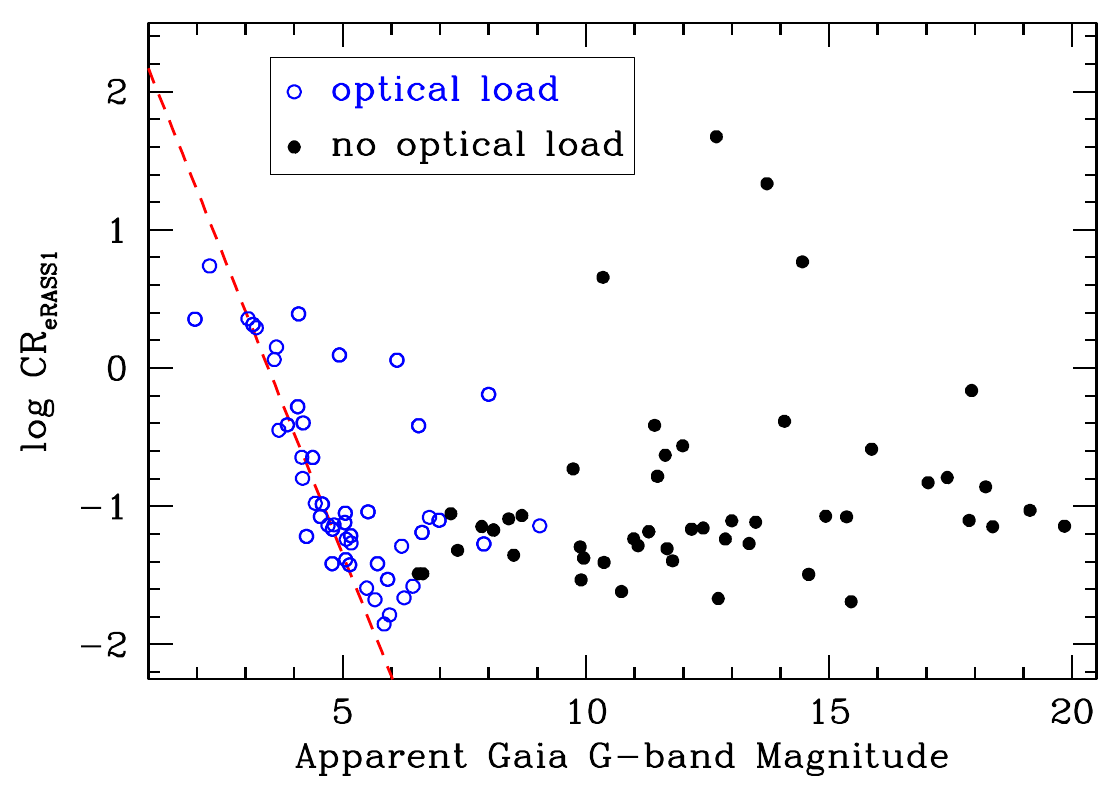}
\caption{
Count rate of eRASS1 ML1 (0.2-2.3 keV) versus {\it Gaia} DR3 $\overline{G}$ magnitudes of AGB stars in eRASS1. 
The blue open symbols correspond to X-ray sources flagged as likely contaminated by optical loading. 
}
\label{fig:cr_g}
\end{figure}

\begin{table*}
\centering
\caption{High confidence X-AGB stars in the eROSITA-DE eRASS1 catalog.}
\label{tbl:agb_lst_ok}
\tiny
\setlength{\tabcolsep}{3pt} 
\renewcommand{\arraystretch}{1.1} 
\begin{tabular}{llrccrrccl}
\hline
IRAS  & Common  & $\overline{G}$~~~~& [$G_{min}$:$G_{max}$] & eRASS1 IAU Name & \multicolumn{1}{c}{Offset}   & \multicolumn{1}{c}{Position 1-$\sigma$}   & 
Counts     &   Count Rate    & Comments \\
Number & Name   & & & & & & & \\             &              &        (mag)      &         (mag)         &                 & \multicolumn{1}{c}{(arcsec)} & \multicolumn{1}{c}{(arcsec)} &           &  (s$^{-1}$)     & \\
\hline
02346$-$6248 & RS\,Hor      &  9.90 &  [7.81:10.68] & 1eRASS J023552.2$-$623501 &  1.4~~~ & 7.0~~~~~~ & $5.6\pm2.6$  & $0.029\pm0.013$ & Mira variable \\
05166$-$2215 & RZ\,Lep      &  7.36 &    $\dots$    & 1eRASS J051846.2$-$221247 &  4.2~~~ & 4.4~~~~~~ &  $8.0\pm3.2$ & $0.048\pm0.019$ & S Star \\
06024$-$8403 & CPD$-$84\,86 &  7.86 &    $\dots$    & 1eRASS J055423.1$-$840321 &  1.7~~~ & 4.6~~~~~~ &   $22\pm5$   & $0.071\pm0.017$ & C-rich/S Star (1) \\
12118$-$5115 & TV\,Cen      &  6.55 &  [6.52:6.80]  & 1eRASS J121431.7$-$513156 &  1.2~~~ & 3.7~~~~~~ &  $6.8\pm3.0$ & $0.033\pm0.014$ & C-rich \\ 
13548$-$3049 & TW\,Cen      &  8.10 &  [5.45:8.30]  & 1eRASS J135743.4$-$310409 &  3.6~~~ & 3.4~~~~~~ & $11.6\pm3.9$ & $0.067\pm0.022$ & O-rich (1) \\
14188$-$6943 & VX\,Cir      & 11.47 &  [9.19:12.00] & 1eRASS J142308.8$-$695737 &  3.1~~~ & 2.0~~~~~~ &   $35\pm6$   & $0.165\pm0.030$ & O-rich, Mira variable \\
16263$-$4910 & ~~$\dots$    & 11.63 & [11.26:11.85] & 1eRASS J163007.2$-$491724 &  4.5~~~ & 2.3~~~~~~ &   $32\pm6$   & $0.235\pm0.045$ & O-rich \\
\hline
\end{tabular}
\tablebib{
(1) \citet{Schmitt2024}.
}
\end{table*}

\begin{table*}
\centering
\caption{Possible X-AGB stars in the eROSITA-DE eRASS1 catalog.}
\label{tbl:agb_lst_poss}
\tiny
\setlength{\tabcolsep}{3pt} 
\renewcommand{\arraystretch}{1.1} 
\begin{tabular}{llrccrrccl}
\hline
IRAS   & Common & $\overline{G}$~~~~& [$G_{min}$:$G_{max}$] & eRASS1 IAU Name & \multicolumn{1}{c}{Offset}   & \multicolumn{1}{c}{Position 1-$\sigma$} & Counts     &   Count Rate    & Comments \\
Number & Name   & & & & & & & \\
             &              &        (mag)      &         (mag)         &                 & \multicolumn{1}{c}{(arcsec)} & \multicolumn{1}{c}{(arcsec)} &           &  (s$^{-1}$)     & \\
\hline
11390$-$7213 & MQ\,Mus      & 10.73 &  [8.72:11.45] & 1eRASS J114119.7$-$723043 &  5.5~~~ & 3.3~~~~~~ & $11.8\pm4.0$ & $0.024\pm0.008$ & Mira variable \\ 
11510$-$6046 & V1245\,Cen   & 14.58 & [13.53:15.38] & 1eRASS J115331.7$-$610325 & 10.4~~~ & 4.9~~~~~~ &   $8.9\pm3.6$ & $0.032\pm0.013$ & C-rich \\ 
~~~~~$\dots$ & CGCS\,3366 & 11.67 & [11.49:11.85] & 1eRASS J131820.6$-$504533 &  7.6~~~ & 4.8~~~~~~ &  $10.2\pm3.7$ & $0.049\pm0.018$ & C-rich \\ 
13494$-$0313 & HD\,120832   &  8.41 &  [8.37:8.48]  & 1eRASS J135201.3$-$032836 & 11.0~~~ & 3.9~~~~~~ & $12.0\pm3.8$ & $0.081\pm0.026$ & S Star, UV source (1) \\ 
16047$-$5449 & V501\,Nor    &  9.95 &  [9.25:10.45] & 1eRASS J160842.5$-$545714 &  8.2~~~ & 4.9~~~~~~ &  $6.2\pm2.8$ & $0.042\pm0.019$ & C-rich \\ 
17023$-$5859 & CH\,Ara      & 10.37 &  [9.25:11.80] & 1eRASS J170647.5$-$590314 &  8.2~~~ & 4.6~~~~~~ & $5.2\pm2.6$ & $0.039\pm0.020$ & Mira variable \\ 
~~~~~$\dots$ & V371\,CrA    & 13.49 & [11.38:14.29] & 1eRASS J181635.8$-$391236 &  6.8~~~ & 4.7~~~~~~ &  $6.3\pm3.0$ & $0.077\pm0.037$ & Mira variable \\ 
\hline
\end{tabular}
\tablebib{
(1) \citet{OG2021}.
}
\end{table*}

\begin{table*}
\centering
\caption{Confused X-AGB stars in the eROSITA-DE eRASS1 catalog.}
\label{tbl:agb_lst_doubt}
\tiny
\setlength{\tabcolsep}{3pt} 
\renewcommand{\arraystretch}{1.1} 
\begin{tabular}{llrccrrccl}
\hline
IRAS   & Common            & $\overline{G}$~~~~& [$G_{min}$:$G_{max}$] & eRASS1 IAU Name & \multicolumn{1}{c}{Offset}   & \multicolumn{1}{c}{Position 1-$\sigma$}   &Counts     &   Count Rate    & Comments \\
Number & Name   & & & & & & & \\             &                       &        (mag)      &         (mag)         &                 & \multicolumn{1}{c}{(arcsec)} & \multicolumn{1}{c}{(arcsec)} &           &  (s$^{-1}$)     & \\
\hline
~~~~~$\dots$ & J132601.58$-$473306.0 & 10.98 & $\dots$ & 1eRASS J132602.0$-$473305 &  4.7~~~ & 3.5~~~~~~ & $11.8\pm3.9$ & $0.058\pm0.019$ & On the outskirts of the \\
\multicolumn{9}{c}{} & Galactic Cluster NGC\,5139 \\
~~~~~$\dots$ & V975\,Oph             & 19.13 & [14.86:19.56] & 1eRASS J174045.3$-$291634 &  2.8~~~ & 3.7~~~~~~ &  $9.6\pm3.4$ & $0.094\pm0.033$ & Mira variable candidate \\ 
\hline
\hline
\end{tabular}
\end{table*}

\begin{table*}
\centering
\caption{Miscellaneous X-ray sources in the eROSITA-DE eRASS1 catalog.}
\label{tbl:other_lst_ok}
\tiny
\setlength{\tabcolsep}{3pt} 
\renewcommand{\arraystretch}{1.1} 
\begin{tabular}{lrccrrccl}
\hline
Common Name  & $\overline{G}$~~~~& [$G_{min}$:$G_{max}$] & eRASS1 IAU Name & \multicolumn{1}{c}{Offset}   & \multicolumn{1}{c}{Position 1-$\sigma$}   & Counts     &   Count Rate    & Comments \\
&        (mag)      &         (mag)         &                 & \multicolumn{1}{c}{(arcsec)} &            &  (s$^{-1}$)     & \\
\hline
BF\,Eri      & 14.45 & [12.79:14.81] & 1eRASS J043929.9$-$043601 &  3.3~~~ & 1.0~~~~~~ &  $792\pm29$  &  $5.88\pm0.22$  & Cataclysmic Variable \\
KT\,Eri      & 14.93 &    $\dots$    & 1eRASS J044754.2$-$101047 &  4.2~~~ & 3.4~~~~~~ & $12.8\pm3.9$ & $0.085\pm0.026$ & Classical Nova \\
J051722.71$-$352156.4 & 10.35 & $\dots$ & 1eRASS J051722.5$-$352159 & 3.7~~~ & 1.0~~~~~~ & $1034\pm33$ & $4.54\pm0.15$ & Low-mass star \\
NQ\,Gem      &  7.22 &  [7.12:7.30]  & 1eRASS J073154.8$+$243026 & 14.2~~~ & 5.1~~~~~~ &  $6.3\pm2.9$ & $0.088\pm0.040$ & Symbiotic Star \\
J082705.08$+$284402.1 & 17.43 & $\dots$ & 1eRASS J082704.9$+$284405 &  3.4~~~ & 3.0~~~~~~ & $12.2\pm3.7$ & $0.162\pm0.048$ & Active dMe in wide binary (1) \\
HD\,100764   &  8.51 &    $\dots$    & 1eRASS J113542.9$-$143536 &  3.3~~~ & 4.5~~~~~~ &  $4.5\pm2.3$ & $0.044\pm0.023$ & Chemically peculiar star, UV source \\ 
RR\,Tel      & 11.41 & [10.27:11.63] & 1eRASS J200418.5$-$554331 &  1.9~~~ & 1.9~~~~~~ &   $36\pm6$   &  $0.39\pm0.07$  & Symbiotic Star \\
\hline
\end{tabular}
\tablebib{
(1) \citet{Farihi+2010}.
}
\end{table*}

\subsection{Cross-correlation of the asymptotic giant branch sample with the eRASS1 catalog}

We searched the eROSITA-DE eRASS1 Main Catalog for X-ray counterparts of AGB stars in the samples presented in Sect.~\ref{subsec:sample}.  
Since sources in the eRASS1 catalog are only to be found in the German "western" hemisphere, the approximately 30,000 AGB stars in these samples actually registered by eRASS1 go down to 13,139 sources. 
Given the typical eRASS1 imaging resolution (HEW$\approx30^{\prime\prime}$) and the possible crowding at the location of AGB stars in the Galactic plane, the search was limited to X-ray sources with a positional coincidence within $20^{\prime\prime}$ of an AGB star.  
This provided a first sample of 130 X-AGB candidates, and they are presented in Figure~\ref{fig:cr_g}. In the figure, we have plotted their count rate in the 0.2-2.3 keV ML1 eROSITA band versus their {\it Gaia} DR3 average $G$ magnitude.  
The plot then distinguishes between sources flagged in the eROSITA-DE eRASS1 catalog as suffering from optical loading (\texttt{FLAG\_OPT} set to 1, blue open circles) and those not flagged (\texttt{FLAG\_OPT} set to 0, black dots).  
Optical loading causes a well-known relation between the observed count rate and the $G$ magnitude \citep[red-dashed line in Fig.~\ref{fig:cr_g},][]{Schmitt2024}, which is followed by stars in our sample brighter than $G < 6$ mag, instead of the canonical $G \leq$ 4.5 mag brightness limit \citep{Merloni+2024}.  
This is probably caused by the large $G-J$ color of AGB stars, but it is also because AGB stars are variable and may have been brighter in the $G$-band at the time they were observed by eROSITA than the average values adopted in this figure. 
Thus, sources brighter than $G \leq 6$ were considered to be affected by optical loading and subsequently discarded regardless of their \texttt{FLAG\_OPT} value.  
On the other hand, sources fainter than $G \geq 6$ but flagged in the eRASS1 catalog as being affected by optical loading (\texttt{FLAG\_OPT} set to 1) were still considered because some of them are listed by \citet{Schmitt2024}.
Finally, sources flagged as being extended were also considered unlikely X-AGB stars and then excluded.
The selection criteria described above provide a final sample of 59 X-AGB candidates in the eROSITA-DE eRASS1 main source catalog.

\section{Data analysis}
\label{sec:analysis}

\subsection{X-ray asymptotic giant branch stars in the eRASS1 catalog}
\label{subsec:cross}

\begin{table*}
\begin{center}
\caption{Results of the spectral analysis performed with XSPEC.}
\tiny
\setlength{\tabcolsep}{\tabcolsep}  
\begin{tabular}{lcccccc} 
\hline
Object & $\chi^{2}_\mathrm{DoF}$ & $N_\mathrm{H}$ & $kT$  & $A$       & $f_\mathrm{X}$  & 
$F_\mathrm{X}$ \\
       &     & (10$^{21}$ cm$^{-2}$)    & (keV) & (10$^{-5}$ cm$^{-5}$) & (erg~cm$^{-2}$~s$^{-1}$)            & (erg~cm$^{-2}$~s$^{-1}$)\\
\hline
CPD$-$84\,86      & 2.7/2.0 = 1.35   & 0.4$^{+0.8}_{-0.8}$  & 2.6$^{+2.0}_{-2.0}$ & 6.6$\pm$1.7 & (5.8$\pm$1.5)$\times10^{-14}$ & (1.1$\pm$0.2)$\times10^{-13}$\\
IRAS~16263$-$4910 & 2.49/3.0 = 0.83 & 0.3$^{+5.0}_{-5.2}$   & 1.9$^{+0.7}_{-0.7}$ & 17.7$\pm$9.3  & (1.95$\pm$1.0)$\times10^{-13}$ & (2.9$\pm$1.5)$\times10^{-13}$  \\
VX Cir            & 8.01/6.0=1.33   & 0.04$^{+1.0}_{-0.04}$ & 1.5$^{+6.7}_{-0.4}$ & 8.8$\pm$3.5  & (1.3$\pm$0.5)$\times10^{-13}$ & (1.6$\pm$0.6)$\times10^{-13}$\\
\hline
HD\,120832        & \dots           &  3.0                  & 2.6                 & 13.4$\pm$5.6 & (6.8$\pm$2.7)$\times10^{-14}$ & (2.2$\pm$0.9)$\times10^{-13}$  \\
\hline
\end{tabular}
\tablefoot{The observed flux ($f_\mathrm{X}$) was computed in the 0.2--2.3 keV energy range, while the intrinsic flux ($F_\mathrm{X}$) was calculated for the 0.3--10.0 keV range. The normalization parameter ($A$) is defined as $A = 10^{-14}\int n_\mathrm{H} n_\mathrm{e} dV / 4 \pi d^{2}$, where $n_\mathrm{H}$ and $n_\mathrm{e}$ are the hydrogen and electron densities, $V$ is the volume of the X-ray-emitting region, and $d$ is the distance.}
\label{tab:spec}
\end{center}
\end{table*}

\begin{figure}
\centering
\includegraphics[bb=21 90 560 465,width=1.0\linewidth]{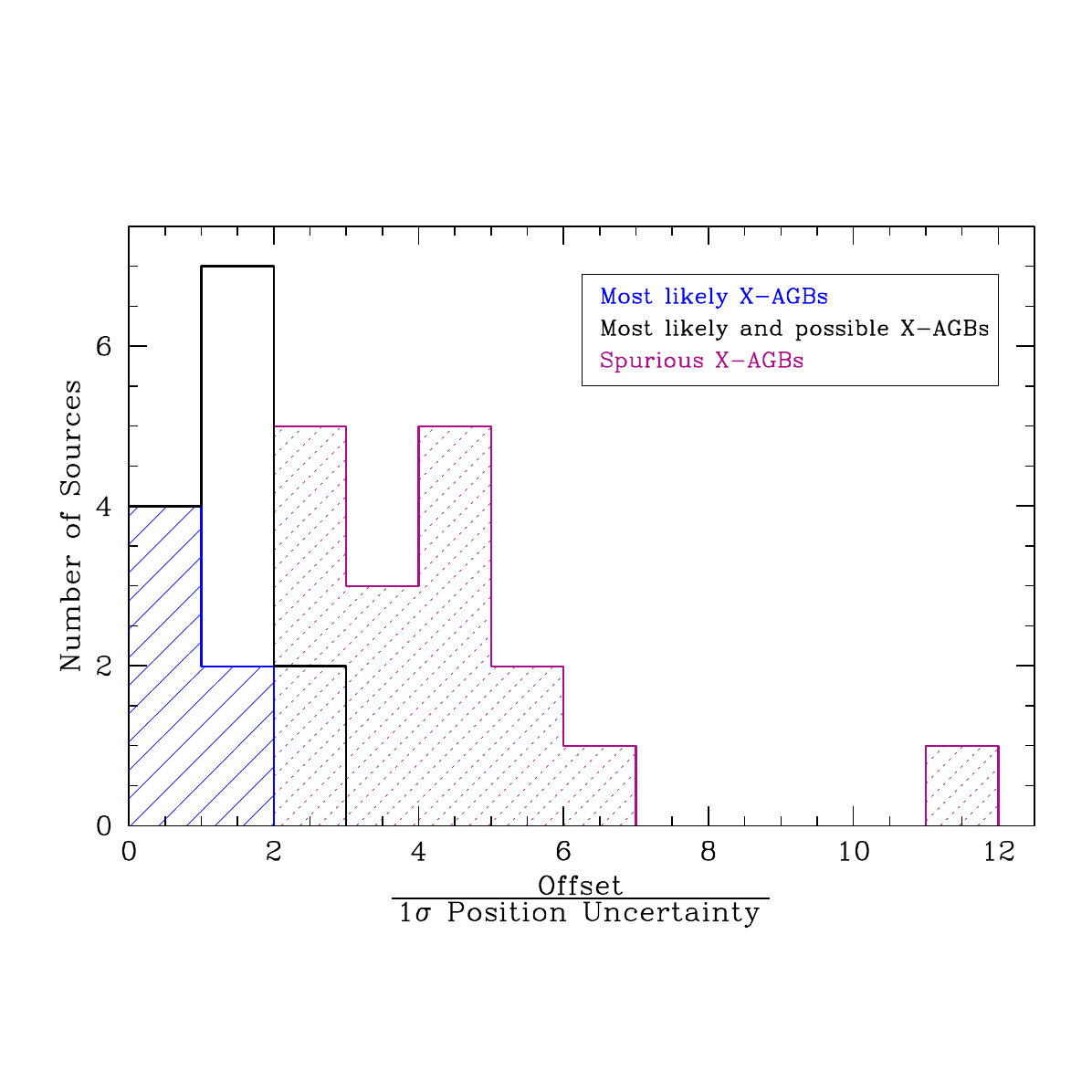}
\caption{
\textbf{}{Histogram of the positional offset between an AGB star and its associated eRASS1 X-ray counterpart relative to the 1-$\sigma$ positional uncertainty of the latter for most likely, possible, and spurious X-AGBs in Tables~\ref{tbl:agb_lst_ok}, \ref{tbl:agb_lst_poss}, and \ref{tbl:agb_lst_ko}, respectively.  
The histogram of the possible X-AGBs (black histogram) was added together with that of the most likely X-AGBs (blue histogram and shade).  
The histogram and shade of the spurious X-AGBs are shown in purple. 
The spurious X-AGBs associated with diffuse X-ray sources or those superimposed on diffuse emission are not included.} 
}
\label{fig:offset_histo}
\end{figure}

\begin{figure*}
\center
\includegraphics[width=0.375\linewidth]{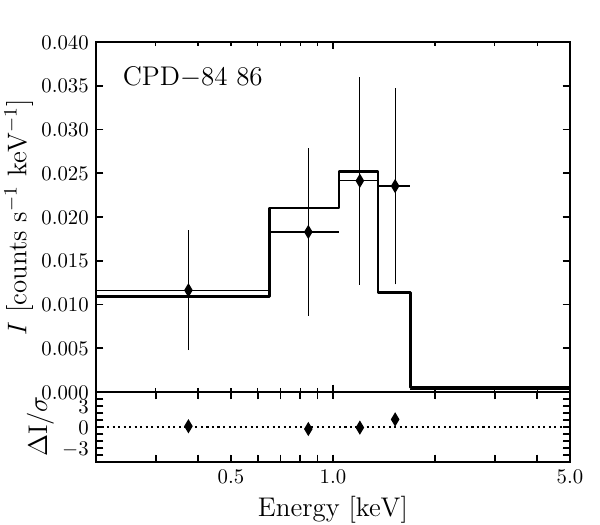}~
\includegraphics[width=0.375\linewidth]{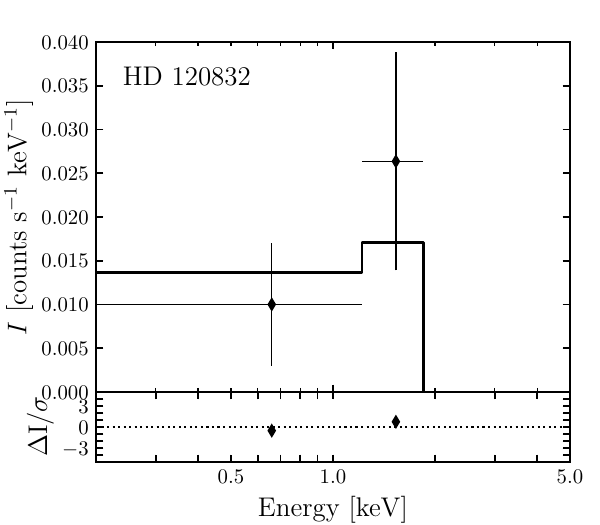}\\
\includegraphics[width=0.375\linewidth]{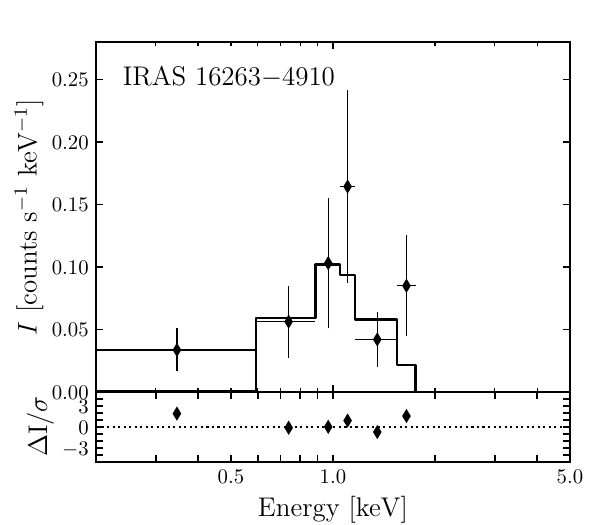}~
\includegraphics[width=0.375\linewidth]{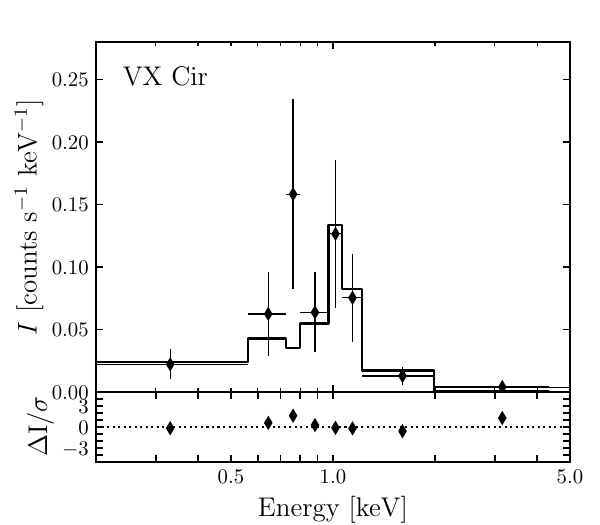}
\caption{Background-subtracted eROSITA spectra of CPD$-$84 86, HD~120832, IRAS~16263$-$4910, and VX~Cir (black diamonds). 
The black solid histograms represent the best fits. 
Fractional residuals are presented in the bottom panels. 
The spectra were binned requesting a minimum of five counts per spectral bin.}
\label{fig:spec}
\end{figure*}

We individually assessed the quality of the association between these 59 X-AGB candidates and their possible X-ray counterparts in the eRASS1 main source catalog.  
We then carefully compared the eROSITA 0.2-2.3 keV X-ray images of these X-AGB candidates to their near-IR 2MASS $K$ images to assess the quality of the positional coincidence between AGB stars and their possible X-ray counterparts.

Sources were then regarded as "high confidence X-AGBs" if there was a positional coincidence within $5^{\prime\prime}$ of the AGB star and its X-ray counterpart, and the latter could not be better associated with any other UV, optical, or near-IR source.\footnote{{\it GALEX}, DSS, and 2MASS images were examined as well.}  
The seven bona fide X-AGBs in the eRASS1 catalog are listed in Table~\ref{tbl:agb_lst_ok}, and their images are presented in Figs.~\ref{fig:IRAS_02346-6248} to \ref{fig:IRAS_16263-4910}.
Table~\ref{tbl:agb_lst_ok} provides their IRAS number and common name (columns \#1 and \#2); their {\it Gaia} DR3 average $\overline{G}$ and minimum and maximum $G$ magnitudes (columns \#3 and \#4); 
the IAU name of their eRASS1 counterpart and spatial offset (columns \#5 and \#6); and the 1-$\sigma$ positional uncertainty, count number, and count rate of the eRASS1 counterpart (columns \#7 to \#9). 
Comments on the sources are given in the last column (column \#10). 
In all cases but one (TW\,Cen), their $G$ magnitudes are larger than 6 mag at any time.

Sources were regarded to be "possible X-AGBs" if there was a positional coincidence between $5^{\prime\prime}$ and $12^{\prime\prime}$ of the AGB star and its X-ray counterpart and the AGB star was the most viable counterpart of the X-ray source at the UV, optical, or near-IR.  
The seven possible X-AGB stars in the eRASS1 catalog are listed in Table~\ref{tbl:agb_lst_poss}, and their images are presented in Figs.~\ref{fig:IRAS_11390-7213} to \ref{fig:V371_CrA}.

The cross-correlation between the X-ray source and a possible AGB counterpart is very dubious for a couple of sources, which are projected against a very crowded field of view.  
These sources, which we considered "confused X-AGBs", are listed in Table~\ref{tbl:agb_lst_doubt}, and their images are presented in Figs.~\ref{fig:J132601.58-473306.0_sidebyside.png} and \ref{fig:V975_Oph}.

Seven of the (presumed to be) AGB stars in our sample with X-ray source counterparts in the eRASS1 catalog have been found, a posteriori, to not be AGB stars according to SIMBAD.  
These seven "miscellaneous X-ray sources" are listed in Table~\ref{tbl:other_lst_ok}. 
We note that the descriptions of the columns in Tables~\ref{tbl:agb_lst_poss}, \ref{tbl:agb_lst_doubt} and \ref{tbl:other_lst_ok} are the same as those in Table~\ref{tbl:agb_lst_ok}.

Finally, spurious X-AGB stars are listed in Table~\ref{tbl:agb_lst_ko}. 
The list includes 11 sources flagged as being affected by optical loading, namely CI Hyi, R\,Hor, U\,Men, U\,Ori, V340\,Car, R\,Gem, Y\,Gem, Y\,Pup, R\,LMi, W\,Vel, and AX\,Sco.  
These sources are generally fainter than $G \geq 6$ mag but brighter than $J \leq 3$, one of the criteria to set a value of one for the \texttt{FLAG\_OPT} flag. 
The list of spurious X-AGB stars also includes 25 extended X-ray sources and/or very unlikely or simply erroneous matches between the AGB star and an X-ray counterpart.  
We note that the average surface density of eRASS1 sources of $\approx3.5\times10^{-6}$ source~arcsec$^2$ implies that 0.0044 sources are expected to be within the 20~arcsec in radius search aperture around each AGB star.  
Therefore a total of $\approx$60 spurious sources are anticipated for the total sample of 13,139 AGB stars registered by eRASS1-DE (i.e.,\ $\approx$45\% chance alignments in the original sample of 130 X-AGB candidates).  
Since this sample of 130 X-AGB candidates was reduced to the 59 that passed the optical loading and brightness criteria, 27 sources can be expected to be the result of changes in alignments, which is notably consistent with the number of spurious sources in Table~\ref{tbl:agb_lst_ko} according to positional criteria.  
Furthermore, the offset between AGB stars and their X-ray counterparts in Tables~\ref{tbl:agb_lst_ok} \textbf{}{and \ref{tbl:agb_lst_poss} is always within 3-$\sigma$} of the X-ray source positional uncertainty provided by the eRASS1 catalog, but \textbf{}{the offset} is larger than 3-$\sigma$ for sources in Table~\ref{tbl:agb_lst_ko}, \textbf{}{with the X-ray source being best associated with a background stellar source in quite a number of cases} (e.g., V677\,Pup, V434\,Vel, V1110\,Cen, and GI\,Lup). 
\textbf{}{
The distribution of the offset between the AGB stars and their X-ray counterparts relative to the 1-$\sigma$ positional uncertainty shown in Figure~\ref{fig:offset_histo} clearly illustrates the inferior quality of the association between AGB stars and spurious X-ray counterparts in Table~\ref{tbl:agb_lst_ko}.  }

To summarize, this search found 14 X-AGBs in the eRASS1 catalog among an input sample of 13,139 AGB stars (with a small fraction of contaminants).  
This implies that one out of 1,000 AGB stars is an X-ray source in the eRASS1 catalog.
Since three of the 14 X-AGBs in Tables~\ref{tbl:agb_lst_ok} and \ref{tbl:agb_lst_poss} had been identified previously in the eRASS1 catalog, namely CPD$-$84\,86 and TW\,Cen \citep{Schmitt2024} and HD\,120832 \citep{OG2021}, the total number of X-AGBs is increased from 36 up to 47, that is, the known sample of X-AGBs is increased by $\simeq30\%$ in this work.

\begin{table}
\centering
\setlength{\tabcolsep}{0.6\tabcolsep} 
\caption{Energy conversion factors (ECFs) for absorbed \texttt{apec} plasma models. 
}
\label{tbl:xray_agb_ecf}
\tiny
\begin{tabular}{crrrrr}
\hline
$N_{\rm H}$ & \multicolumn{5}{c}{$T_{\rm X}$} \\
\cmidrule(lr){2-6}
 ($10^{22} {\rm cm}^{-2}$) &  3 MK &  10 MK &  30 MK &  50 MK &  90 MK \\
\hline
0.03 & 9.164e+11 & 1.115e+12 & 9.204e+11 & 8.979e+11 & 8.839e+11 \\
0.10 & 5.836e+11 & 8.709e+11 & 7.244e+11 & 7.062e+11 & 6.974e+11 \\
0.30 & 2.088e+11 & 4.767e+11 & {\bf 4.535e+11} & 4.450e+11 & 4.438e+11 \\
1.00 & 1.829e+10 & 9.847e+10 & 1.711e+11 & 1.766e+11 & 1.830e+11 \\
2.00 & 2.765e+09 & 2.566e+10 & 6.954e+10 & 7.591e+10 & 8.164e+10 \\
\hline
\end{tabular}
\tablefoot{
The values of ECFs, computed according to Eq.~\ref{eq:ecf}, are provided in units of centimeters squared by erg. 
The value of the ECF for the "standard X-AGB spectral model" ($N_{\rm H} = 3\times10^{21}$~cm$^{-2}$, $T_{\rm X} = 30$~MK) is shown in boldface.}
\end{table}

\subsection{Spectral analysis}

The cross-correlation of the eRASS1 catalog with the sample of AGB stars identified seven high confidence X-AGBs (Table~\ref{tbl:agb_lst_ok}) and another seven possible X-AGBs (Table~\ref{tbl:agb_lst_poss}). 
The eRASS1 catalog includes source and background spectra only for the few sources with a sufficient count number, namely CPD$-$84\,86 (IRAS\,06024$-$8403), VX\,Cir (IRAS\,14188-6943), and IRAS\,16263$-$4910 in Tab.~\ref{tbl:agb_lst_ok} and HD\,120832 (IRAS\,13494$-$0313) in Tab.~\ref{tbl:agb_lst_poss}.  
These spectra were obtained by combining multiple exposures of the seven modules on board {\it eROSITA}. 
The corresponding RMF and ARF calibration files are provided by the eRASS1 catalog as well. 
The background-subtracted spectra of CPD$-$84\,86, HD\,120832, VX~Cir, and IRAS\,16263$-$4910 are presented in Fig.~\ref{fig:spec}.  
All spectra peak at $\approx1$~keV, and there is no emission above 2 keV.  
These spectral properties discard nonthermal emission and rather hint at plasma emission with a high temperature.

The spectral analysis of CPD$-$84\,86, IRAS\,16263$-$4910, and VX\,Cir was performed using the X-ray spectral fitting package \citep[XSPEC;][]{Arnaud1996}. 
A grouped file was created by using the \texttt{grppha} task of HEASoft\footnote{\url{https://heasarc.gsfc.nasa.gov/docs/software/heasoft/}} requesting five counts per bin.
Models adopting a \texttt{tbabs} absorption model in conjunction with an optically thin \texttt{apec} emission plasma model with solar abundances were attempted in the energy range with detected emission, mostly below 2 keV. 
The best-fit hydrogen column density $N_\mathrm{H}$ and plasma temperature $T_\mathrm{X}$ listed in Table~\ref{tab:spec} are consistent with those reported for X-AGBs by \citet{OG2021}. 
Slightly different abundances might be needed in order to improve the fit, but such a detailed model is not well justified given the small count number.

\begin{table}
\centering
\setlength{\tabcolsep}{\tabcolsep} 
\caption{Flux correction unitless factors (FCFs) of the eRASS1 catalog for absorbed \texttt{apec} plasma models. 
}
\label{tbl:xray_agb_cfc}
\tiny
\begin{tabular}{crrrrr}
\hline
$N_{\rm H}$ & \multicolumn{5}{c}{$T_{\rm X}$} \\
\cmidrule(lr){2-6}
($10^{22} {\rm cm}^{-2}$) &  3 MK &  10 MK &  30 MK &  50 MK &  90 MK \\
\hline
0.03 &    1.172 &    0.963 &    1.167 &    1.196 &    1.215 \\
0.10 &    1.840 &    1.233 &    1.483 &    1.521 &    1.540 \\
0.30 &    5.144 &    2.253 &    {\bf 2.368} &    2.413 &    2.420 \\
1.00 &   58.735 &   10.907 &    6.277 &    6.082 &    5.870 \\
2.00 &  388.400 &   41.858 &   15.445 &   14.149 &   13.155 \\
\hline
\end{tabular}
\tablefoot{
FCFs are computed according to Eq.~\ref{eq:fcf}. 
The value of FCF for the "standard X-AGB spectral model" ($N_{\rm H} = 3\times10^{21}$~cm$^{-2}$, $T_{\rm X} = 30$~MK) is shown in boldface. }
\end{table}

\subsection{X-ray flux and luminosity of X-AGBs in eRASS1}

The majority of X-AGBs and potential X-AGBs in Tables~\ref{tbl:agb_lst_ok} and \ref{tbl:agb_lst_poss}, respectively, do not have sufficient eRASS1 counts for spectral analysis. 
For those sources, the eRASS1 catalog provides estimates of the X-ray flux, applying to the observed count rates an energy conversion factor (ECF) derived from an emission model consisting of an absorbed power-law with a slope of 2.0 and an absorbing column density $N_{\rm H} = 3\times10^{20}$ cm$^{-2}$ typical of Galactic absorption \citep{Brunner+2022,Merloni+2024} 
This emission model, which is suitable for AGNs \citep{Liu+2022}, is obviously not satisfactory for X-AGBs, whose X-ray emission is best described by an absorbed optically thin plasma emission model, such as the \texttt{apec} model, with plasma temperatures in the range of 4 MK to 117 MK \citep{Sahai2015,OG2021}.   
Therefore, appropriate ECFs were computed for a grid of an absorbed \texttt{apec} plasma emission model from count rates in the 0.2$-$2.3 keV eROSITA ML1 band to observed X-ray fluxes (Table~\ref{tbl:xray_agb_ecf}) using the following equation: 
\begin{equation}
\label{eq:ecf}
    f_{\rm X} = {\rm ML\_RATE\_1} / {\rm ECF}. 
\end{equation}
The typical hydrogen column densities and plasma temperatures of X-AGBs \citep[Table~\ref{tab:spec} and][]{OG2021} are found to be $3\times10^{21}$ cm$^{-2}$ and 30 MK, respectively.  
This ``standard X-AGB spectral model'' results in the ``central'' value in Table~\ref{tbl:xray_agb_ecf} (ECF$_0\approx4.5\times10^{11}$ cm$^2$~erg$^{-1}$; shown in boldface), which we used in this work for the energy conversion.  

For comparison, the eRASS1 catalog flux correction factors (FCF) to \texttt{apec} plasma model fluxes 
\begin{equation}
\label{eq:fcf}
    f_{\rm X} = {\rm ML\_FLUX\_1} / {\rm FCF} 
\end{equation}
are shown in Table~\ref{tbl:xray_agb_cfc}. 
The standard X-AGB spectral model results in the central value in Table~\ref{tbl:xray_agb_cfc} (FCF$_0=2.368$), which implies that the X-ray flux of the X-AGBs derived above is $\approx2.4$ times larger than those listed in the eRASS1 catalog.

We note that most X-AGBs are actually found to have $N_{\rm H}$ values in the range from $1\times10^{21}$ to $1\times10^{22}$ cm$^{-2}$ and $T_{\rm X}$ in the range from 10 to 50 MK.  
The ECFs for these values of $N_{\rm H}$ and $T_{\rm X}$, which encircle the central value in Table~\ref{tbl:xray_agb_ecf}, reveal very likely excursions for the value ECF$_0$ adopted here that can amount up to $\simeq$90\%.  
Hence, for the likely extremes of the values of $N_{\rm H}$ and $T_{\rm X}$, particularly for high absorption columns and low plasma temperatures, the value of the ECF may depart notably from the one adopted here.

\begin{table*}
\centering
\caption{
Observed X-ray flux (0.2$-$2.3 keV) and intrinsic luminosity (0.3$-$10.0 keV) of eRASS1 X-AGBs. }
\label{tbl:xray_agb_lst}
\tiny
\setlength{\tabcolsep}{5pt} 
\renewcommand{\arraystretch}{1.1} 
\begin{tabular}{llrrrrrcr}
\hline
IRAS Number & Common Name & \multicolumn{1}{c}{$f_{\rm X}$}       & Distance &    \multicolumn{1}{c}{$L_{\rm X}$} & \multicolumn{1}{c}{$L/L_{\odot}$} & \multicolumn{1}{c}{$L_{\rm X}/L$} & \multicolumn{1}{c}{$\log (L_{\rm X}/L)$} & \multicolumn{1}{c}{$T_{\rm X}$} \\
           &   &  ($10^{-14}$ erg~cm$^{-2}$~s$^{-1}$)  & (pc)~~~~ & ($10^{30}$ erg~s$^{-1}$) & ($\times 10^3$)& ($\times 10^{-6}$) & & \multicolumn{1}{c}{(MK)} \\
\hline
\multicolumn{8}{c}{High confidence X-AGB stars} \\
\hline
02346$-$6248 & RS\,Hor      &   $6.4\pm3.0$~~~~~~~~ & 1710$\pm$160~~  &  $64\pm32$~~  &  1.7~~ &  9.8~~ & $-5.0$ & $\dots$ \\
05166$-$2215 & RZ\,Lep      &  $10.6\pm4.2$~~~~~~~~ & 1070$\pm$60~~~~ &  $41\pm17$~~  &  1.8~~ &  6.0~~ & $-5.2$ & $\dots$ \\
06024$-$8403 & CPD$-$84\,86 &   $5.8\pm1.5$~~~~~~~~ &  793$\pm$20~~~~ &  $8.3\pm1.6$~ &  1.2~~ &  1.7~~ & $-5.8$ & $30\pm23$ \\
12118$-$5115 & TV\,Cen      &   $7.3\pm3.2$~~~~~~~~ &  849$\pm$17~~~~ &  $18\pm8$~~~~ &  0.59  &  7.9~~ & $-5.1$ & $\dots$ \\
13548$-$3049 & TW\,Cen      &  $15\pm5$~~~~~~~~~~~  &  770$\pm$70~~~~ &  $30\pm12$~~  &  3.7~~ &  2.1~~ & $-5.7$ & $\dots$ \\
14188$-$6943 & VX\,Cir      &  $13\pm5$~~~~~~~~~~~  & 2450$\pm$430~~  & $115\pm60$~~  & 25.4~~ &  1.2~~ & $-5.9$ & $17^{+78}_{-5}$ \\
16263$-$4910 & ~~~~$\dots$  & $20\pm10$~~~~~~~~~    & 1550$\pm$260~~  & $160\pm100$   &  1.6~~ & 26.9~~ & $-4.6$ & $22\pm8$ \\
\hline
\multicolumn{8}{c}{Possible X-AGB stars} \\
\hline
11390$-$7213 & MQ\,Mus       &  $5.3\pm1.8$~~~~~~~~ & 2040$\pm$360~~  &  $75\pm37$~~ &     7.3~~  &    2.7~~ & $-5.6$ & $\dots$ \\
11510$-$6046 & V1245\,Cen    &  $7.1\pm2.9$~~~~~~~~ & $\dots$~~~~~    & $\dots$~~~~  & $\dots$~ & $\dots$~~~ & $\dots$ & $\dots$ \\
~~~~~~~~~$ \dots$ & 
             CGCS\,3366 & $10.8\pm4.0$~~~~~~~~ & 5000$\pm$700~~  & $910\pm420$  &     0.8~~ &   290~~~~ & $-3.5$ & $\dots$ \\
13494$-$0313 & HD\,120832    & $18\pm6$~~~~~~~~~~~  &  853$\pm$34~~~~ &  $45\pm15$~~ &     0.7~~ &   17~~~~ & $-4.8$ & $\dots$ \\
16047$-$5449 & V501\,Nor     &  $9.3\pm4.2$~~~~~~~~ & 2020$\pm$190~~  & $130\pm60$~~ &     7.2~~ &    4.7~~ & $-5.3$ & $\dots$ \\
17023$-$5859 & CH\,Ara       &  $8.6\pm4.4$~~~~~~~~ & 3700$\pm$1000   & $390\pm290$  &     7.8~~ &   13~~~~ & $-4.9$ & $\dots$ \\
~~~~~~~~~$\dots$ & V371\,CrA & $17\pm8$~~~~~~~~~~~  & $\dots$~~~~~    & $\dots$~~~~  & $\dots$~ & $\dots$~~~ & $\dots$ & $\dots$ \\
\hline
\end{tabular}
\tablefoot{As described in the text, the relative uncertainty in $L/L_{\odot}$ is smaller to $\approx^{+74\%}_{-42\%}$.}
\end{table*}

Based on the ECF ratio described above, the observed X-ray flux and intrinsic X-ray luminosity of the AGB stars whose spectral fit was not possible are listed in Table \ref{tbl:xray_agb_lst}. 
The standard X-AGB spectral model was also used to derive the observed X-ray flux and intrinsic X-ray luminosity of HD\,120832 (also known as IRAS\,13494$-$0313), whose photon-starved spectrum made performing a reliable spectral fit impossible. 
Otherwise, the ECFs that can be derived from Eq.~\ref{eq:ecf} using the measured count rates and X-ray fluxes derived from the spectral fits of CPD--84\,86, IRAS\,16263$-$4910, and VX\,Cir have values of $\simeq1.2\times10^{12}$ cm$^2$~erg$^{-1}$, which are consistent with those expected for sources with low hydrogen column densities and temperatures in the range of 10 to 30 MK (Table~\ref{tbl:xray_agb_ecf} ).

The stellar luminosity ($L$) in Table~\ref{tbl:xray_agb_lst} was obtained from {\it Gaia} parallaxes and near-IR magnitudes using the ($J-K$) $\times$  BC$_K$ relationship derived by \citet{Whitelock2000,Whitelock2006}. The uncertainty in luminosity depends on the error in the stellar distance ($D$) and the amplitude of variability ($\Delta M$). Table \ref{tbl:xray_agb_lst} lists the parallax distances along with their uncertainties. 
All objects closer than 1 kpc show ${\sigma}_D/D < 10$\%, while those beyond 1 kpc exhibit ${\sigma}_D/D < 18$\% (except for CH\,Ara, ${\sigma}_D/D = 27$\%). 
As a result, the bolometric magnitudes would be uncertain by up to $0.21^{\rm m}$ and $0.36^{\rm m}$, respectively ($0.52^{\rm m}$ for CH\,Ara). 
For AGB Mira and semi-regular (SR) variables, the amplitude of the bolometric magnitude is strongly correlated with the $K$-band amplitude \citep[$\Delta M_{\rm bol}=1.210 \Delta K$,][]{Whitelock2000}. 
Unfortunately, there is no similar correlation between $\Delta M_{\rm bol}$ and $\Delta V$ (or $\Delta m_v$), and only one star in our sample (V1245\,Cen) has been monitored in the $K$-band \citep[$\Delta K=0.78$, implying $\Delta M_{\rm bol}=0.94$,][]{Whitelock2006}. 
Consequently, except for this star, we cannot individually determine the effect of the variability on the stellar luminosity. 
However, in a sample consisting mainly of 92 Mira-type variables found in the {\it HIPPARCOS} catalog, \citet{Whitelock2000} observed $0.0 < \Delta M_{\rm bol} < 1.2$, which corresponds to a maximum variation in luminosity of $\Delta L$ $^{+74\%}_{-42\%}$ or $-0.24 < \Delta (\log L) < +0.24$. Therefore, the effect caused by the error in the parallax is generally less important than the variation of luminosity caused by pulsation, especially for the high-amplitude Mira-type variables.

Table~\ref{tbl:xray_agb_lst} shows the properties of the X-AGBs found in our cross-correlation of the samples of AGB stars described in Sect.~\ref{subsec:sample} and the eRASS1 catalog. 
Eleven are completely new discoveries, whereas three have been reported previously, including CPD$-$84\,86 and TW\,Cen \citep{Schmitt2024} and HD\,120832 \citep{OG2021}.  
Most X-AGBs in Table~\ref{tbl:xray_agb_lst} show $L_X/L_{\rm bol}$ larger than the previous upper limit for this ratio established by \citet{OG2021}: $L_{\rm X}/L_{\rm bol} \lesssim 3 \times 10^{-6}$. 
Notably, the AGB star IRAS\,16263$-$4910 exhibits an exceptionally high relative luminosity ($L_{\rm X}/L_{\rm bol}=2.69 \times 10^{-5}$), much higher than most X-AGBs and X-SySts considered by \citet{OG2021}.

\section{Discussion}
\label{sec:discussion}

Since the first discoveries of X-ray emission associated with AGB stars, the origin of their high-energy radiation has been attributed to various mechanisms that also operate in X-SySts and/or cataclysmic variables (CVs), such as shell burning at the surface of the WD \citep{Orio2007} and/or shocks produced by the winds of both components \citep{Muerset1997}. 
The X-ray emission of confirmed SySts (i.e., showing typical SySts spectral features, such as the He~{\sc ii}~$\lambda$4686 emission line and O~{\sc vi}~$\lambda$6825 Raman band) 
can certainly be associated with a degenerate companion \citep{Orio2007}.   
The scenario involving accretion onto a degenerate companion, however, is found to be inconsistent in most X-AGBs due to their low X-ray luminosity. 
In these cases, a main-sequence star accreting mass from the primary AGB seems more plausible given the observed X-ray luminosity \citep{KS2004}. 

\begin{figure}
\centering
\includegraphics[width=1.0\linewidth]{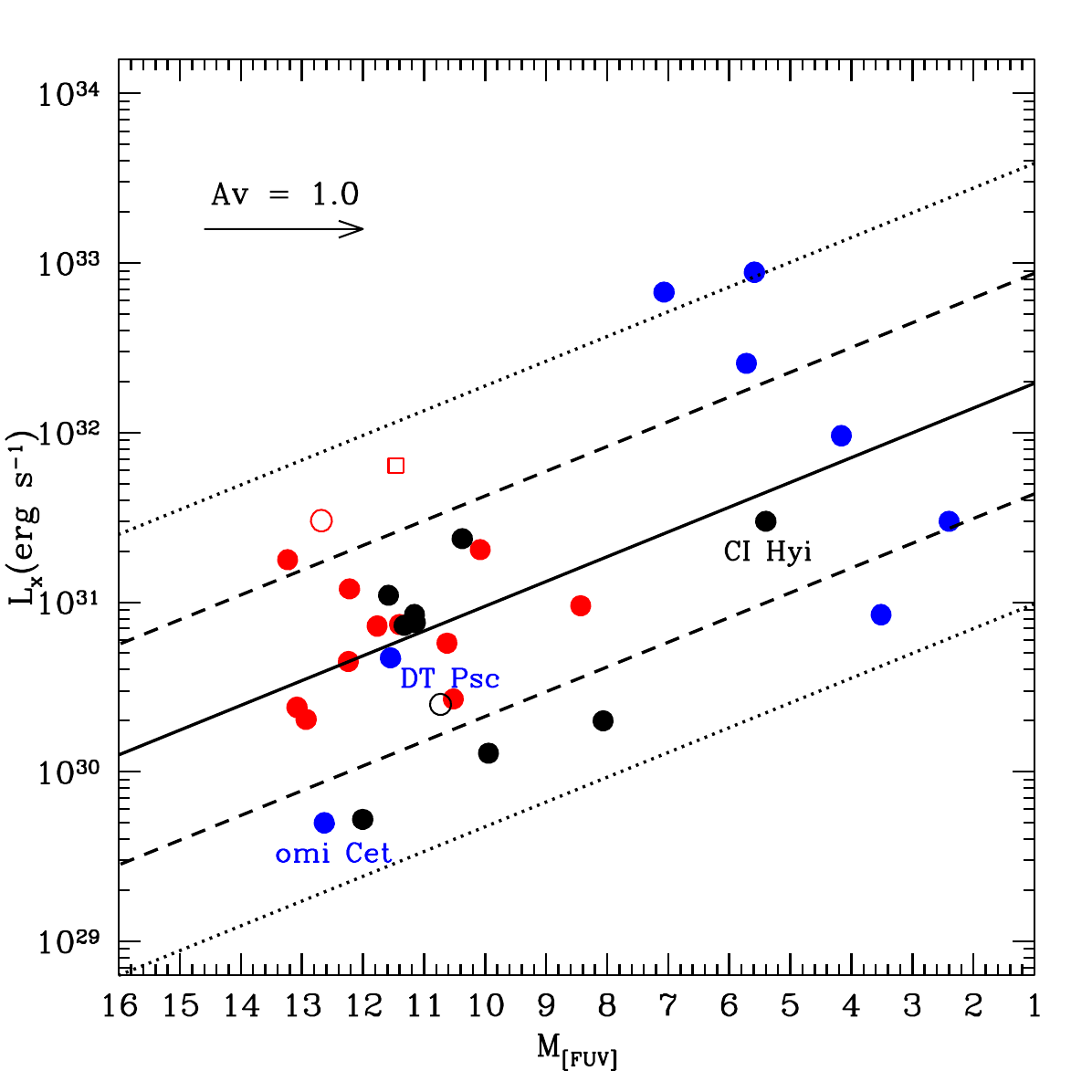}
\caption{{\it GALEX} [FUV] absolute magnitude (M$_{\rm [FUV]}$) versus the X-ray luminosity of AGB and SySts. The X-ray AGB stars discovered by eROSITA are plotted with red symbols \citep[squares = this work; circles =][]{Schmitt2024}; X-AGBs discovered by other X-ray surveys are shown with black circles; and SySts are shown with blue circles. The open symbols represent upper limits for M$_{\rm [FUV]}$, calculated at the {\it GALEX} detection threshold according to \citet{Montez2017}. The arrow represents the far-UV extinction corresponding to one visual magnitude, assuming $A_{\rm [FUV]}/A_{\rm V}=2.6$ \citep{Cardelli1989}. 
The solid line shows a least-squared fit (see Eq.~\ref{eq:label}) to all detected sources, while the dashed and dotted lines represent the 1-$\sigma$ and 2-$\sigma$ confidence levels, respectively. 
}
\label{lxlfuv}
\end{figure}

The objects classified as ``X-ray AGB stars'' (or simply X-AGBs) in the literature and in the present study generally do not show the emission lines characteristic of X-SySts \citep[but see the case of the Fe K-shell 6.4 and 6.7 keV emission lines in Y\,Gem,][]{OG2021}. 
Thus, the nature of their X-ray emission is still debated. 
For X-AGBs with main-sequence or WD companions, the AGB star would generally outshine the companion in the infrared and visual spectral regions. 
However, in the far-UV, the companion can outshine the AGB star.
Therefore a comparison between their UV and X-ray properties may shed light on the mechanisms that produce high-energy photons in AGB stars.

Far-UV excess in AGBs has been detected in the {\it GALEX} bands and in {\it IUE} spectra \citep[e.g.][]{OG2016,Sahai2022}.
Figure~\ref{lxlfuv} presents the $L_\mathrm{X}$--M$_{\rm [FUV]}$ diagram for X-SySts and X-AGBs for which far-UV data are available in the {\it GALEX} archive. 
The list of X-SySts was taken from the New Online database of Symbiotic Variables\footnote{\url{https://sirrah.troja.mff.cuni.cz/~merc/nodsv/}} \citep{Merc2019}; included in the plot are the Galactic X-SySts observed in X-rays during quiescence, whereas X-AGBs were taken from various references in the literature \citep{Hunsch1998,KS2004,Ramstedt2012,Sahai2015,OG2021,Schmitt2024}. 
We note that some sources formerly known as X-AGBs have been suggested to be X-SySts \citep[e.g., Y\,Gem;][]{Yu+2022}, but Fig.~\ref{lxlfuv} does not include those unless their SySt nature has been confirmed \citep{Merc2019}.

X-AGBs are generally distributed in the $L_\mathrm{X}$--M$_{\rm [FUV]}$ diagram as an extension of the X-SySts branch toward lower far-UV and X-ray luminosities.  
Although X-AGBs can be better disentangled from X-SySts by their M$_{\rm [FUV]}$, the X-SySts DT\,Psc and $o$\,Cet (aka Mira) exhibit very low X-ray and far-UV luminosity, consistent with those of X-AGBs. 
These two X-ray faint X-SySts are actually ``weakly SySts,'' where the red giant and the WD are in a wide orbit.  
Thus, the abnormally low X-ray luminosity of $o$ Cet, $L_\mathrm{X} = 5 \times 10^{29}$ erg~s$^{-1}$ initially attributed to a low-mass main-sequence companion \citep{JH1984,KS2004} had to be reassessed when its WD nature was confirmed. 
The low X-ray luminosity of $o$ Cet was accordingly ascribed to a low accretion rate onto the WD on a wide orbit. 
Meanwhile, DT\,Psc, a Tc-poor S3/2-type AGB star showing a highly variable X-ray emission \citep{Jorissen1996}, has a relatively low far-UV flux that does not seem to support a high temperature WD companion.  
Monitoring of its radial velocity resulted in an orbital period 
of 4596~yr and $e = 0.18$ \citep{Jorissen2019}, which implies a large separation between the two components except at periastron. 
On the other hand, the X-AGB CI\,Hyi exhibits a very high far-UV and X-ray luminosity\footnote{
\citet{Sahai2015} determined an X-ray luminosity of $3.0\times10^{31}$ erg~s$^{-1}$ for a plasma with temperature $7.4\times10^7$~K using XMM-Newton data.} comparable to the bright X-SySts.

Apart from these exceptional cases, a correlation between far-UV and X-ray luminosity is found when X-AGBs and X-SySts are considered altogether. 
Generally, objects brighter in the far-UV have higher X-ray luminosity. 
The linear regression fit to all sources in the $L_\mathrm{X}$--M$_\mathrm{[FUV]}$ diagram,  
\begin{equation}
\log(L_\mathrm{X}) = 32.44 -(0.146 \cdot  \mathrm{M}_\mathrm{[FUV]}),
    \label{eq:label}
\end{equation}
\noindent is plotted in Fig.~\ref{lxlfuv}.  
The fit has a low regression coefficient, $\approx0.34$, and a non-negligible standard deviation, $\sigma = 0.65$, though it implies only a mild correlation. 

\begin{figure}
\centering
\includegraphics[width=1.0\linewidth]{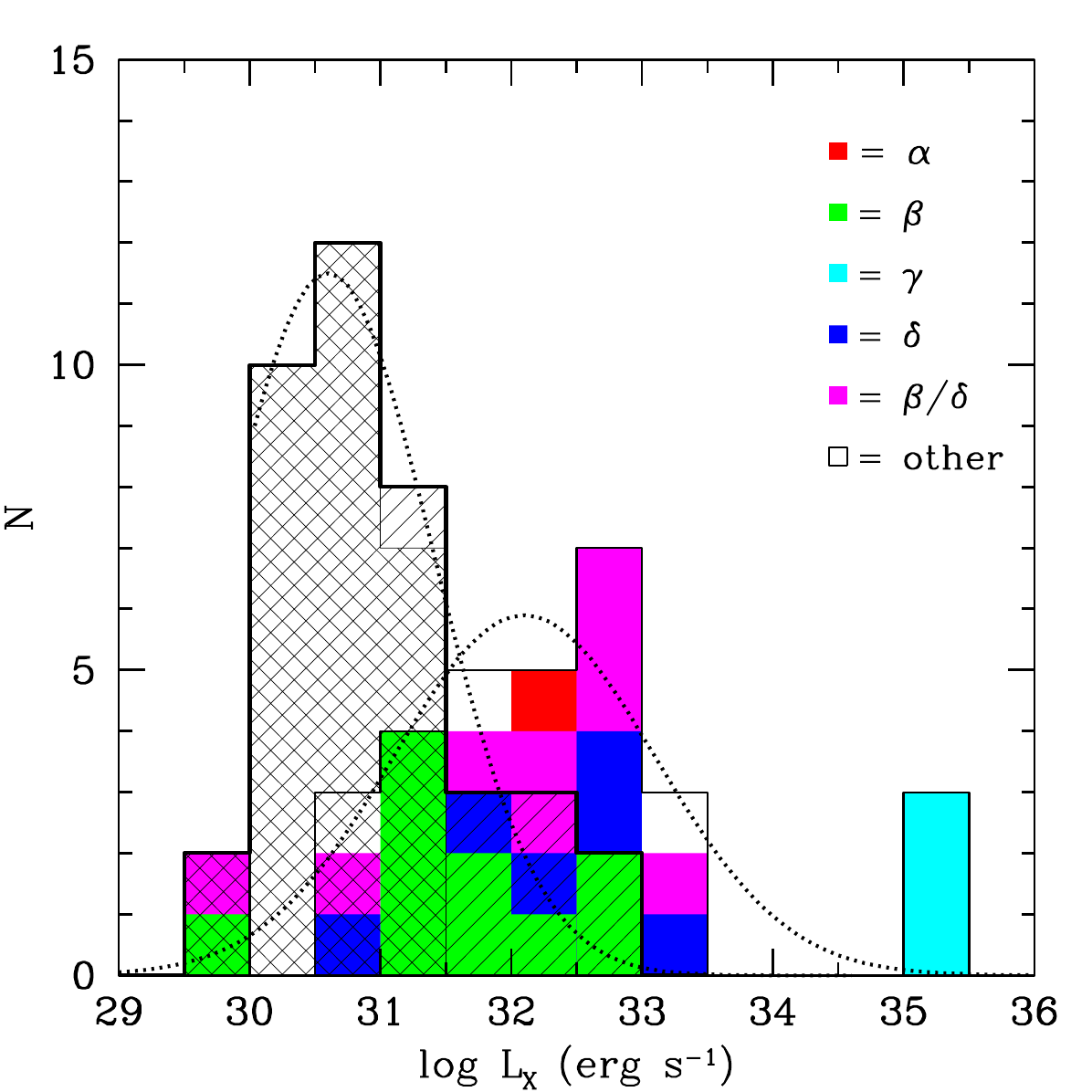}
\caption{
Histogram of the X-ray luminosity of AGB stars and SySts. Previously known X-AGB stars are shown with black crossed-dashed symbols, newly found X-AGBs are shown with black diagonal-dashed symbols. The X-SySts are shown with colors, according to the classification of their X-ray spectra, as labeled.
The dotted and solid histograms are the best least-squared Gaussian fits to the X-AGBs and X-SySts, respectively. }
\label{histo_X}
\end{figure}

The X-ray luminosity distributions of X-AGBs and X-SySts are shown more comprehensively in Fig.~\ref{histo_X} because many objects, those not having a {\it GALEX} detection in the far-UV band, are absent in Fig.~\ref{lxlfuv}. 
The X-SySts in the figure are highlighted by different colors depending on their X-ray spectral type \citep{Muerset1997,Luna2013}, from the softest sources ($\alpha$-type) to those emitting in the hard X-ray regime ($\delta$-type). 
For comparison, we also plot those of the $\gamma$-type, but it must be noted that their compact object is accepted to be a neutron star \citep[see][]{Merc2019} 
and their spectra have a nonthermal origin.

The X-ray luminosity distributions of X-AGBs and X-SySts in Fig.~\ref{histo_X} can be fitted with Gaussian curves to each population corresponding to averaged values of $\log(L_\mathrm{X}/\mathrm{erg~s}^{-1})$=30.6$\pm$0.8 for X-AGBs and $\log(L_\mathrm{X}/\mathrm{erg~s}^{-1})$=32.1$\pm$1.0 for X-SySts. 
The X-ray luminosities of AGBs below $\log (L_{\rm X}/{\rm erg s}^{-1}) < 30.0$ were not considered in the fit because of the strong selection effect that prevents the detection of distant, low-luminosity X-ray sources.
Actually, most X-AGBs found in the present work have an X-ray luminosity more than one standard deviation above the average $L_\mathrm{X}$ (i.e., $\log (L_{\rm X}/{\rm erg~s}^{-1})> 31.4$). 
This bias is probably caused by the low sensitivity of the eROSITA eRASS1 survey when compared with previous pointed XMM-Newton or Chandra observations.

A two-sample Kolmogorov-Smirnov test to check whether the samples of X-AGBs and X-SySts were drawn from the same population resulted in a $p$-value of about 0.0015, which is much smaller than the typical threshold of 0.05 to reject the null hypothesis. 
A two-sample Anderson-Darling test also resulted in a similar conclusion.  
Although the two samples are not drawn from the same population, the overlap of the X-ray distributions of X-AGBs and X-SySts is notorious (apart from the extremely high X-ray luminosity of $\gamma$-type X-SySts), 
as both exhibit X-ray luminosity as low as a few times $10^{29}$ erg~s$^{-1}$. 
Therefore, it is not possible to determine whether an X-ray source is an X-AGB or an X-SySt within the interval $10^{29.5} < L_\mathrm{X}$ (erg s$^{-1}$) $< 10^{33.0}$ and based solely on its X-ray luminosity.

Some X-AGBs showing $L_\mathrm{X} \gtrsim 10^{32}$ erg s$^{-1}$ and/or M$_{\rm [FUV]} \lesssim 8.0$ can actually very likely be SySts not recognized so far. 
There are four objects detected in the present study that exceed this limit of luminosity, but none of them were detected and/or observed by {\it GALEX} in the far-UV: the ``high-confidence'' X-AGBs VX\,Cir and IRAS\,16263$-$4910 and the ``possible'' X-AGBs V501\,Nor and CH\,Ara (Table \ref{tbl:xray_agb_lst}). 
As for CGCS\,3366, it shows $L_\mathrm{X} \simeq 10^{33}$ erg s$^{-1}$, but this figure is dubious due to its uncertain distance, the largest among all objects in this study.
The plasma temperatures of VX\,Cir and IRAS\,16263$-$4910 are 17.4 and 22.0 MK, respectively, similar to the typical coronal temperatures of A- to K-type giants \citep[$\approx 1.0 \times 10^7$ K,][]{Schmitt1990}, but their X-ray luminosity ($L_\mathrm{X}=(1.15 \pm 0.60) \times 10^{32}$ and $(1.60 \pm 1.00) \times 10^{32}$, respectively) are near the maximal value of normal A- to K-type giants \citep[$L_\mathrm{X} < 10^{32}$ erg s$^{-1}$,][]{Schmitt1990}. 
On the other hand, the wide interval of temperatures observed among X-SySts \citep[4\,MK $< T_{\rm plasma} <$ 130\,MK;][]{Nunez2014,Nunez2016} does not allow one to classify a source as an X-AGB or an X-SySt based solely on their plasma temperature.

The luminosity ratio $L_\mathrm{X}/L$ of the eRASS1 X-AGBs, mostly in the range $10^{-6}$ to $2\times10^{-5}$, is similar to that of X-SySts with an AGB star \citep[see the left panel of figure 12 in][]{OG2021}. 
This fact in conjunction with the compatible X-ray temperature and X-ray luminosity within the interval $29.5 < \log (L_\mathrm{X}/\mathrm{erg~s}^{-1}) < 33.0$ of X-AGBs and X-SySts noted above implies that their X-ray emission mechanisms may be similar. 
The initial X-ray spectral classification of X-SySts proposed different emission mechanisms
that could generate X-rays in SySts: 
$\alpha$-type have very soft spectra ($E < 0.4$ keV), usually associated with thermonuclear burning on the WD surface \citep{Orio2007}; 
$\beta$-type have harder spectra, peaking at $\sim$0.8 keV and initially attributed to the collision of the stellar winds of both components or to jets \citep{Luna2013,Muerset1997}; and 
$\delta$-type have highly absorbed harder spectra ($E$>4 keV) with Fe emission lines in the 6.0–7.0 keV energy range, most probably generated in the boundary layer between the accretion disk and the WD. 
Finally the spectra of $\beta$/$\delta$-type spectra have similar properties to the $\delta$-type but with a considerable contribution in the 0.3--4.0 keV energy range. 
There is convincing evidence, however, that most X-SySts can be interpreted within a unified scenario consisting of an accretion disk with an X-ray-emitting boundary layer and the additional occurrence of jets, where the different X-SySts types depend on the viewing angle and the different accretion conditions \citep[see][]{Toala2024}.

Under this scenario, the X-ray-emitting plasma in X-SySts of $\beta$-, $\delta$-, and $\beta$/$\delta$-type spectra is expected to at least be produced at the inner boundary layer of the accretion disk, 
with the possible contribution of soft X-ray emission from jets.
The temperature of the plasma there depends on the efficiency of the accretion process: The formation of plasma with temperature of $\sim$10$^{5}$~K is expected for highly efficient accreting sources, while low efficiency accreting ones are expected to produce plasma with temperatures of $\sim$10$^{8}$~K \citep{PS1979,PR1985}. 
If accretion is indeed the same mechanism producing the X-ray emission from X-AGB stars, one can estimate accretion rates from their X-ray luminosities. 
For the typical X-ray luminosity of X-SySts derived above of $\log(L_\mathrm{X}/\mathrm{erg~s}^{-1})$=32.0$\pm$0.8, a standard accretion rate of 1.3$\times10^{15}$~g~s$^{-1}$ ($\approx2\times10^{-11}$~M$_\odot$~yr$^{-1}$) can be estimated using the expression

\begin{equation}
    \dot{M}_\mathrm{acc} \approx \frac{L_\mathrm{X}}{G}\frac{R}{M}, 
\end{equation}

\noindent assuming that all compact companions in X-SySts are WDs with a radius of 0.015~R$_\odot$ and typical masses of 0.6~M$_\odot$. 
If the X-ray emission in X-AGBs has the same origin and efficiency, similar accretion rates imply solar-like (F-K type) stars as companions in X-AGB systems for their typical $\log(L_\mathrm{X}/\mathrm{erg~s}^{-1})$=30.6$\pm$0.8 X-ray luminosity.

\section{Conclusions and summary}
\label{sec:conclusions}

We have presented a search for X-ray counterparts of AGB stars in the eROSITA-DE eRASS1, the first eROSITA data release. 
Seven X-AGBs are confirmed and another seven are identified as possible X-AGBs, increasing the known sample of X-AGBs from 36 up to 47 (three out of these fourteen eRASS1 X-AGBs had been identified in a previous X-ray search). 
This confirms that X-ray emission is ordinary among AGB stars and that the eROSITA sky survey is very useful for detecting X-AGBs despite the fact that the majority of the new X-AGB stars found in the present work are more than one standard deviation above the average X-ray luminosity of all X-AGBs known to date.
The first eROSITA sky survey is thus only sensitive to the brightest X-AGBs, but future higher-sensitivity data releases will be capable of detecting an increasing number of X-AGB stars, especially those showing $L_X < 10^{30}$ erg s$^{-1}$.

The X-ray emission of the eRASS1 X-AGBs with sufficient quality spectrum can be described by an optically thin thermal plasma emission model.  
The best-fit plasma temperatures and column densities of the X-AGBs presented here are similar to those of previously identified X-AGBs. 
The X-ray luminosity of the eRASS1 X-AGBs with a low count number has been derived using the count rate to energy conversion factor for a ``standard X-AGB spectral model'' with $N_{\rm H} = 3\times10^{21}$ cm$^{-2}$ and $T_{\rm X} = 30$ MK.

In this work, we extended the comparison of the X-ray and far-UV properties of X-AGBs with those of X-SySts presented by \citet{OG2021}. 
After analyzing the far-UV and X-ray fluxes of all X-AGBs and X-SySts known to date, we confirm the relationship between the far-UV and X-ray luminosity formerly described by those authors.  
Besides, X-SySts tend to have higher far-UV and X-ray luminosity than X-AGBs, even though there is a notable overlap in the $29.5 < \log (L_X)$(erg s$^{-1}$)$ < 33.0$ luminosity range.  
Most X-SySts within this luminosity range have $\beta$-, $\delta$-, or $\beta/\delta$-type X-ray spectra, suggesting that X-AGBs may share the same emission mechanisms (i.e., colliding winds and a boundary layer between an accretion disk and a companion, or a combination of both).  
Whereas the typical X-ray luminosity $L_\mathrm{X} \approx 10^{32}$ erg~s $^{-1}$ of X-SySts requires a WD accretor, the lower typical X-ray luminosity of X-AGBs, $L_\mathrm{X} \approx 5\times10^{30}$ erg~s $^{-1}$, implies a main-sequence solar-like F-K accretor.

The X-AGBs VX\,Cir, IRAS\,16263$-$4910, V501\,Nor, and CH\,Ara exhibit the highest X-ray luminosities in the sample, similar to the maximum values observed among normal A- to K-giant stars. 
Consequently, they are considered potential SySts, and perhaps spectroscopic observations can eventually confirm this suspicion.

Future releases of eRASS will certainly discover new X-AGBs. 
Since the X-ray properties of $\beta$, $\delta$, and $\beta/\delta$ type SySts are well-known to evolve fast in time, a similar behavior can be expected in X-AGBs, and therefore, this behavior should be monitored in time.  
Thus future releases of eRASS and/or devoted X-ray observations of X-AGBs may reveal an intrinsic variability in these sources as well.

\begin{acknowledgements}

M.A.G.\ acknowledges financial support from grants CEX2021-001131-S funded by MCIN/AEI/10.13039/501100011033 and PID2022-142925NB-I00  from the Spanish Ministerio de Ciencia, Innovaci\'on y Universidades (MCIU) cofunded with FEDER funds.
J.A.T. thanks Direcci\'{o}n General de Asuntos del Personal Acad\'{e}mico (DGAPA) of the Universidad Nacional Aut\'{o}noma de M\'{e}xico (UNAM, Mexico) project IA102324. R.O. acknowledges the S\~ao Paulo Research Foundation (FAPESP) for its
financial support under grant \#2023/05298-0. \\
This work is based on data from eROSITA, the soft X-ray instrument aboard SRG, a joint Russian-German science mission supported by the Russian Space Agency (Roskosmos), in the interests of the Russian Academy of Sciences represented by its Space Research Institute (IKI), and the Deutsches Zentrum für Luft- und Raumfahrt (DLR). The SRG spacecraft was built by Lavochkin Association (NPOL) and its subcontractors, and is operated by NPOL with support from the Max Planck Institute for Extraterrestrial Physics (MPE). The development and construction of the eROSITA X-ray instrument was led by MPE, with contributions from the Dr. Karl Remeis Observatory Bamberg \& ECAP (FAU Erlangen-Nuernberg), the University of Hamburg Observatory, the Leibniz Institute for Astrophysics Potsdam (AIP), and the Institute for Astronomy and Astrophysics of the University of Tübingen, with the support of DLR and the Max Planck Society. The Argelander Institute for Astronomy of the University of Bonn and the Ludwig Maximilians Universität Munich also participated in the science preparation for eROSITA. 
This publication makes also use of data products from the Two Micron All Sky Survey, which is a joint project of the University of Massachusetts and the Infrared Processing and Analysis Center/California Institute of Technology, funded by the National Aeronautics and Space Administration and the National Science Foundation. 
Finally, this work has made extensive use of the NASA's Astrophysics Data System (ADS). 

\end{acknowledgements}

\bibliographystyle{aa} 


\begin{appendix}

\section{eROSITA X-ray and 2MASS near-infrared images of high confidence X-AGB stars}
\label{app:appendix.img.ok}

In this appendix we present a comparison between X-ray and near-IR images of the sources classified as very likely detection in the eROSITA first data release. 

\begin{figure}
\centering
\includegraphics[width=1.\linewidth]{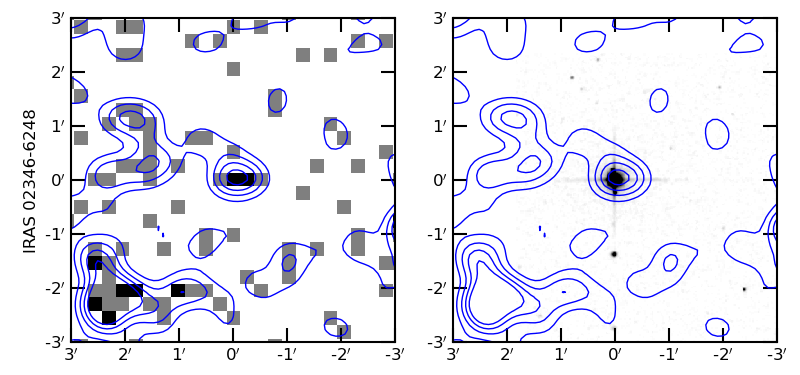}
\caption{Images of eROSITA eRASS1 0.2-2.3 keV (left) and 2MASS $K$ (right) overlaid with X-ray contours of IRAS\,02346$-$6248 (also known as RS\,Hor). The contours have been derived from a resampled and Gaussian-smoothed X-ray image with levels at 30\%, 50\%, 70\%, and 90\% of the count peak in the central $2^{\prime}\times2^{\prime}$ region.
}
\label{fig:IRAS_02346-6248}
\end{figure}

\begin{figure}
\vspace*{-0.5cm}
\centering
\includegraphics[width=1.\linewidth]{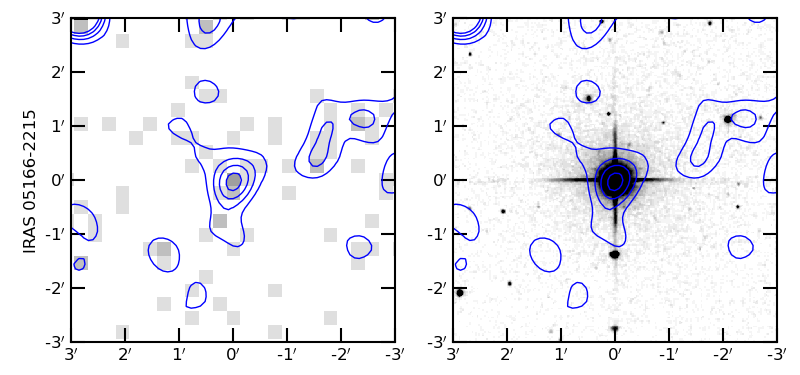}
\caption{Same as Fig.~\ref{fig:IRAS_02346-6248} but for IRAS\,05166$-$2215 (also known as RZ\,Lep).}
\label{fig:IRAS_05166-2215}
\end{figure}

\begin{figure}
\vspace*{-0.1cm}
\centering
\includegraphics[width=1.\linewidth]{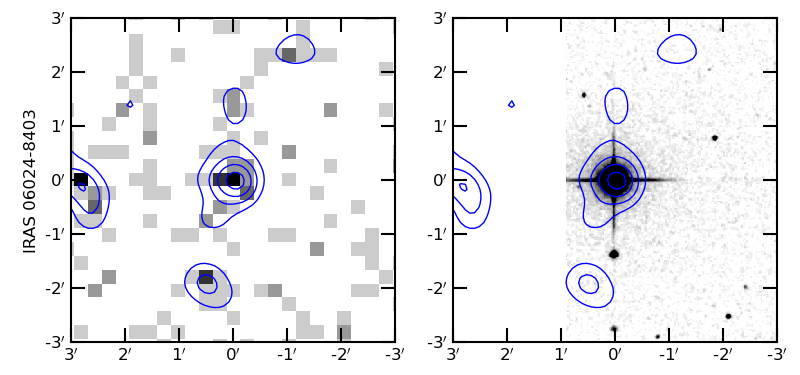}
\caption{Same as Fig.~\ref{fig:IRAS_02346-6248} but for IRAS\,06024$-$8403 (also known as CPD--84\,86).}
\label{fig:IRAS_06024-8403}
\end{figure}

\begin{figure}
\vspace*{-0.1cm}
\centering
\includegraphics[width=1.\linewidth]{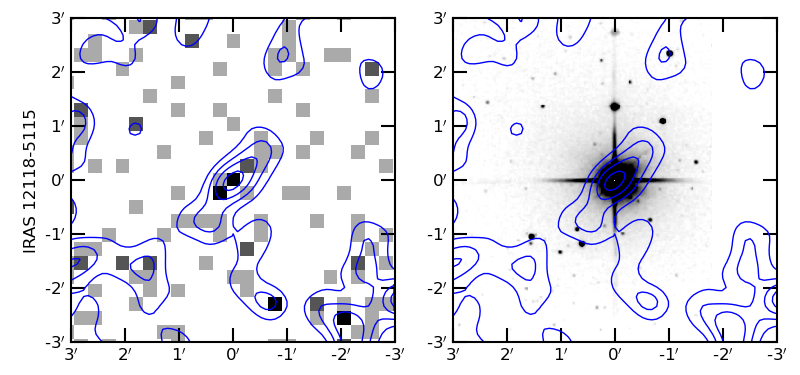}
\caption{Same as Fig.~\ref{fig:IRAS_02346-6248} but for IRAS\,12118$-$5115 (also known as TV\,Cen).}
\label{fig:IRAS_12118-5115}
\end{figure}

\begin{figure}
\centering
\includegraphics[width=1.\linewidth]{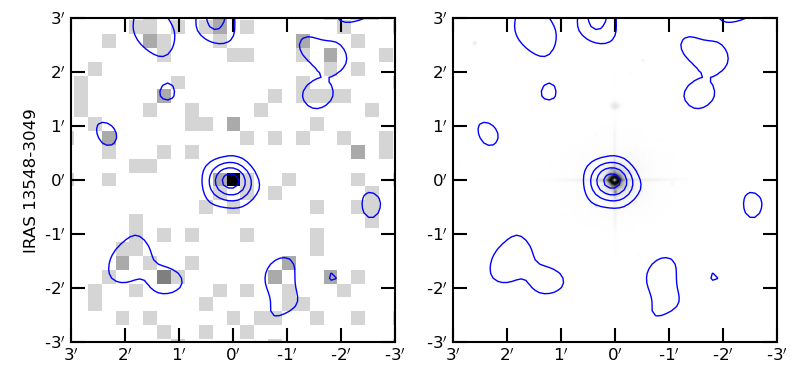}
\caption{Same as Fig.~\ref{fig:IRAS_02346-6248} but for IRAS\,13548$-$3049 (also known as TW\,Cen).}
\label{fig:IRAS_13548-3049}
\end{figure}

\begin{figure}
\centering
\includegraphics[width=1.\linewidth]{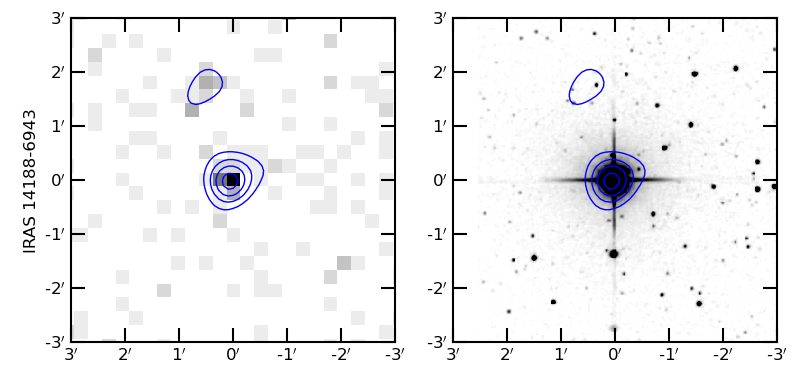}
\caption{Same as Fig.~\ref{fig:IRAS_02346-6248} but for IRAS\,14188$-$6943 (also known as VX\,Cir).}
\label{fig:IRAS_14188-6943}
\end{figure}

\begin{figure}
\centering
\includegraphics[width=1\linewidth]{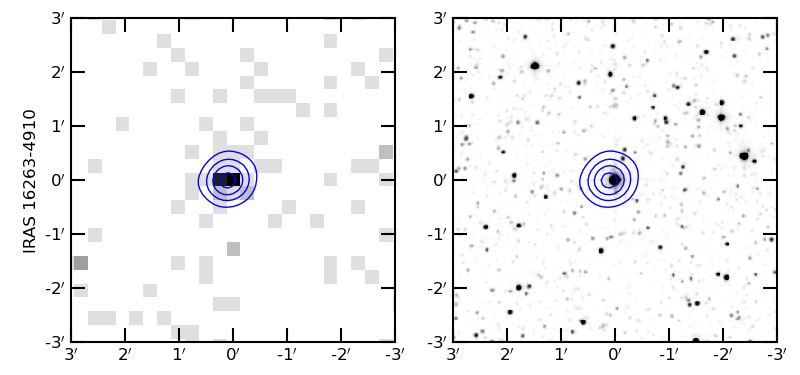}
\caption{Same as Fig.~\ref{fig:IRAS_02346-6248} but for IRAS\,16263$-$4910.}
\label{fig:IRAS_16263-4910}
\end{figure}

\clearpage

\section{eROSITA X-ray and 2MASS near-infrared images of possible X-AGB stars}
\label{app:appendix.img.possible}

In this appendix we present X-ray and near-IR images of the sources defined as possible X-ray detections of AGB stars in the eROSITA first data release.

\begin{figure}
\centering
\includegraphics[width=1.\linewidth]{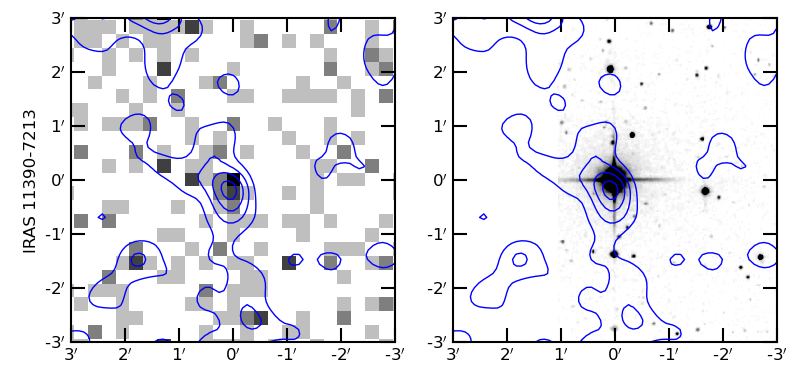}
\caption{Same as Fig.~\ref{fig:IRAS_02346-6248} but for IRAS\,11390$-$7213 (also known as MQ\,Mus).}
\label{fig:IRAS_11390-7213}
\end{figure}

\begin{figure}
\centering
\includegraphics[width=1.\linewidth]{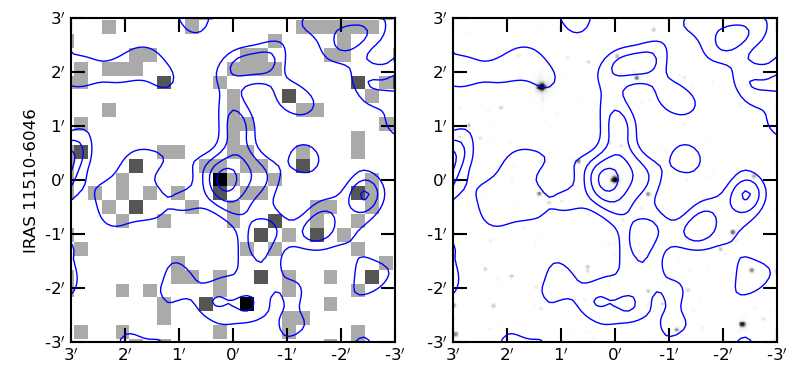}
\caption{Same as Fig.~\ref{fig:IRAS_02346-6248} but for IRAS\,11510$-$6046 (also known as V1245\,Cen).}
\label{fig:IRAS_11510-6046}
\end{figure}

\begin{figure}
\center
\includegraphics[width=1.\linewidth]{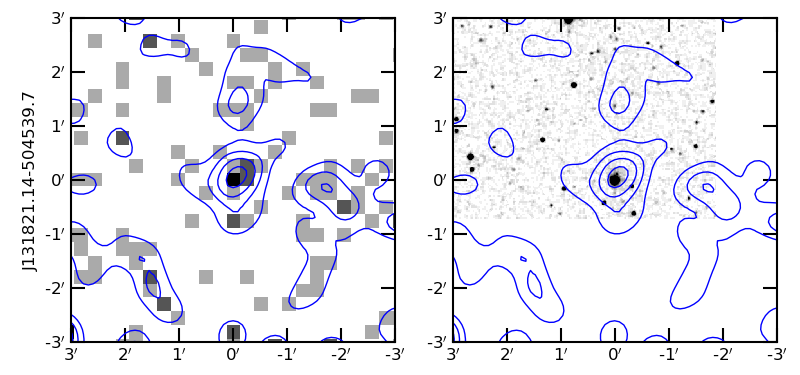}
\caption{Same as Fig.~\ref{fig:IRAS_02346-6248} but for J131821.14$-$504539.7 (also known as CGCS 3366).}
\label{fig:J131821.14-504539.7_sidebyside.png}
\end{figure}

\begin{figure}
\centering
\includegraphics[width=1.\linewidth]{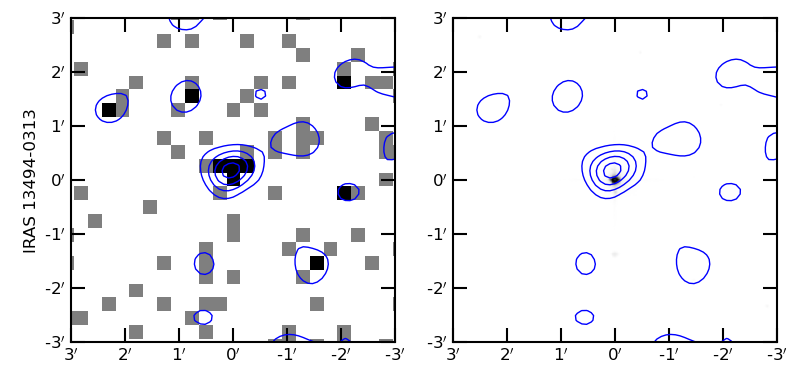}
\caption{Same as Fig.~\ref{fig:IRAS_02346-6248} but for IRAS\,13494$-$0313 (also known as HD\,120832).}
\label{fig:IRAS_13494-0313}
\end{figure}

\begin{figure}
\centering
\includegraphics[width=1.\linewidth]{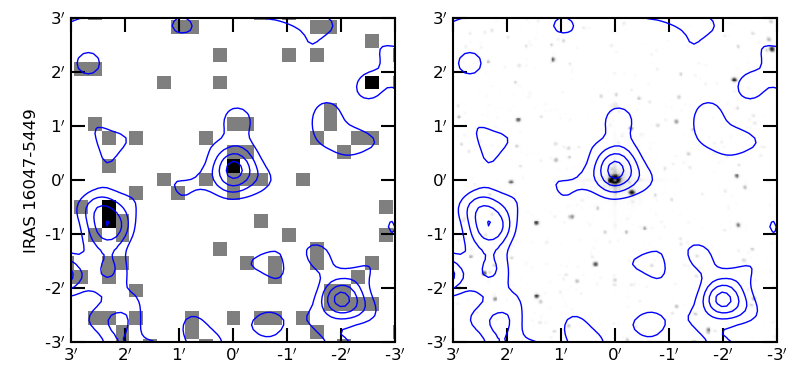}
\caption{Same as Fig.~\ref{fig:IRAS_02346-6248} but for IRAS\,16047$-$5449 (also known as V501\,Nor).}
\label{fig:IRAS_16047-5449}
\end{figure}

\begin{figure}
\centering
\includegraphics[width=1.\linewidth]{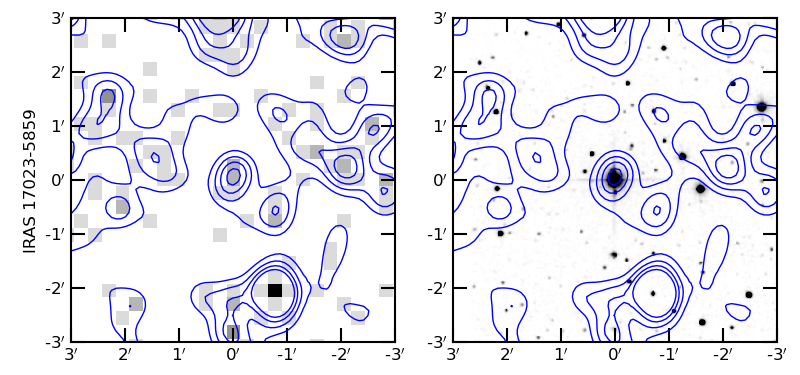}
\caption{Same as Fig.~\ref{fig:IRAS_02346-6248} but for IRAS\,17023$-$5859 (also known as CH\,Ara).}
\label{fig:IRAS_17023-5859}
\end{figure}

\begin{figure}
\centering
\includegraphics[width=1.\linewidth]{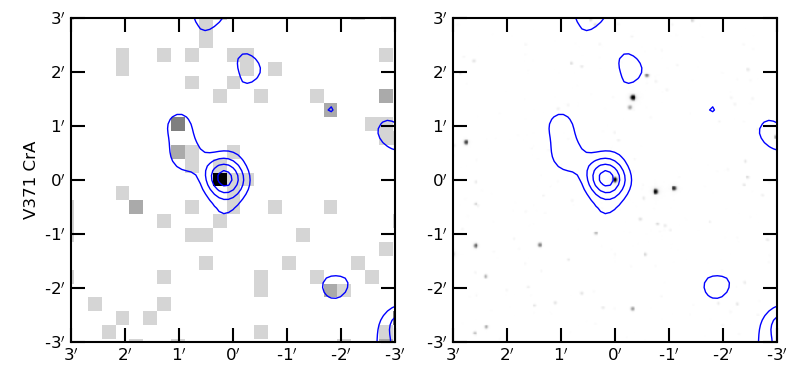}
\caption{Same as Fig.~\ref{fig:IRAS_02346-6248} but for V371\,CrA. }
\label{fig:V371_CrA}
\end{figure}

\clearpage

\section{eROSITA X-ray and 2MASS near-infrared images of confused X-AGB stars}
\label{app:appendix.img.confused}

\begin{figure}
\center
\includegraphics[width=1.\linewidth]{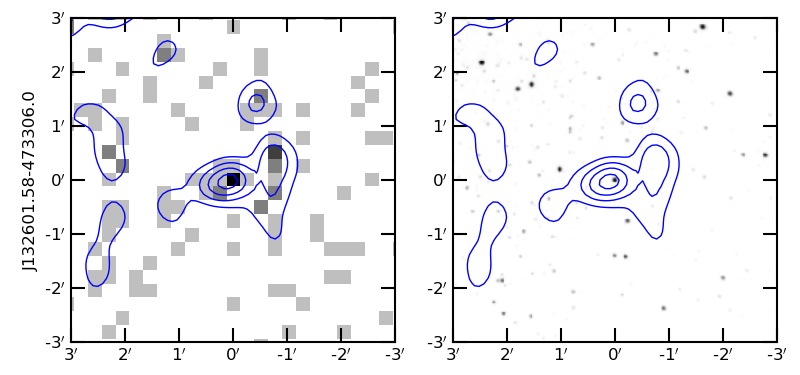}
\caption{Same as Fig.~\ref{fig:IRAS_02346-6248} but for J132601.58$-$473306.0 (aka CGCS 3382).}
\label{fig:J132601.58-473306.0_sidebyside.png}
\end{figure}

\begin{figure}
\centering
\includegraphics[width=1.\linewidth]{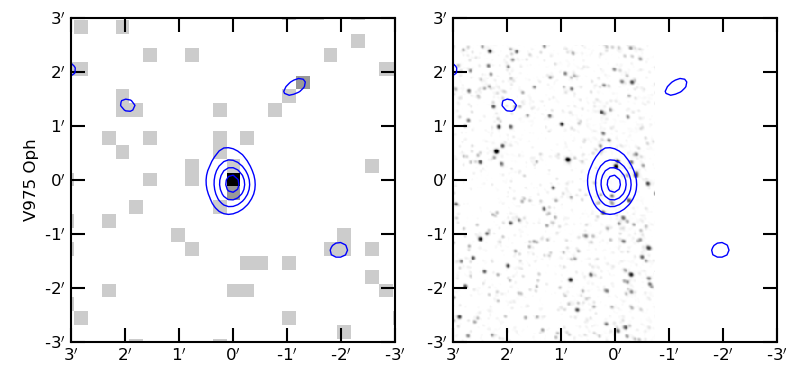}
\caption{Same as Fig.~\ref{fig:IRAS_02346-6248} but for V975\,Oph.}
\label{fig:V975_Oph}
\end{figure}

\clearpage

\section{Spurious X-AGB stars in the eROSITA eRASS1 catalog} 
\label{app:appendix.tab.ko}

\begin{table*}
\centering
\caption{Spurious X-AGB stars in the eROSITA eRASS1 catalog.}
\label{tbl:agb_lst_ko}
\tiny
\setlength{\tabcolsep}{3pt} 
\renewcommand{\arraystretch}{1.1} 
\begin{tabular}{llrccrrl}
\hline
IRAS Number  & Common Name  & $\overline{G}$~~~~&  [$G_{min}$:$G_{max}$] & eRASS1 IAU Name &  \multicolumn{1}{c}{Offset}  & \multicolumn{1}{c}{Position 1-$\sigma$}  & Comments \\
             &              &        (mag)      &          (mag)         &                 & \multicolumn{1}{c}{(arcsec)} & \multicolumn{1}{c}{(arcsec)} & \\
\hline
02110$-$7143 & CI\,Hyi      &  6.78 &  [6.59:6.96]  & 1eRASS J021149.9$-$712904 &  5.8~~~ & 3.5~~~~~~ & Optical loading, LP variable (1,2) \\
02522$-$5005 & R\,Hor       &  6.12 &    $\dots$    & 1eRASS J025353.0$-$495323 &  3.1~~~ & 1.0~~~~~~ & Optical loading, Mira variable (1) \\
04140$-$8158 & U\,Men       &  8.00 &  [5.55:8.41]  & 1eRASS J040935.5$-$815117 &  0.4~~~ & 1.2~~~~~~ & Optical loading (1) \\
05528+2010 & U\,Ori         &  6.63 &  [3.94:6.98]  & 1eRASS J055549.0$+$201028 &  2.7~~~ & 4.8~~~~~~ & Optical loading, Mira variable \\
06397$-$5223 & V340\,Car & 6.26 & [5.78:6.52] & 1eRASS J064053.5$-$522602 & 12.7~~~ & 4.1~~~~~~ & Optical loading, LP variable \\
07043$+$2246 & R\,Gem       &  9.05 &  [5.52:9.31]  & 1eRASS J070721.7$+$224210 & 7.5~~~ & 4.0~~~~~~ & Optical loading, S Star \\
07382$+$2032 & Y\,Gem       &  6.56 &  [6.22:6.75]  & 1eRASS J074108.4$+$202539 &  4.4~~~ & 2.0~~~~~~ & Optical loading, likely Symbiotic Star (2,3) \\ 
07333$-$2217 & ~~$\dots$    & $\dots$ & $\dots$     & 1eRASS J073527.6$-$222339 & 15.3~~~ & 4.4~~~~~~ & C-rich star embedded in reflection nebula, \\
\multicolumn{7}{c}{} & no X-IR match \\
07560$-$4543 & V677\,Pup    & 11.07 & [10.26:11.58] & 1eRASS J075740.3$-$455123 & 18.9~~~ & 3.9~~~~~~ &  Mira variable, background X-ray source \\
08107$-$3459 & Y\,Pup & 6.44 & [6.07:6.66] & 1eRASS J081237.1$-$350848 & 3.4~~~ & 5.2~~~~~~ & Optical loading, LP variable \\
08470$-$4542 & ~~$\dots$    & 18.22 & [17.18:18.92] & 1eRASS J084844.6$-$455345 & 10.1~~~ & 3.9~~~~~~ & Carbon Star, high background emission \\
~~~~~~~~$\dots$ & V930\,Car & 12.68 & $\dots$ & 1eRASS J091202.1$-$645204 & 15.9~~~ & 1.4~~~~~~ & Projected at the core of the GC NGC\,2808 \\
~~~~~~~~$\dots$ & J093912.09$-$573608.8 & 13.36 & $\dots$ & 1eRASS J093910.3$-$573607 & 13.7~~~ & 2.8~~~~~~ & Wrong coordinates, should be 093908.7$-$573607, \\
\multicolumn{7}{c}{} & no X-IR match \\
09425+3444 & R\,LMi         &  6.98 &  [4.89:7.98]  & 1eRASS J094534.5$+$343043 &  3.7~~~ & 3.6~~~~~~ & Optical loading (1) \\
10133$-$5413 & W\,Vel       &  7.90 &  [5.85:8.48]  & 1eRASS J101515.5$-$542845 &  7.1~~~ & 3.1~~~~~~ & Optical loading, Mira variable \\
10449$-$4339 & V434\,Vel    & 11.29 & [10.32:11.71  & 1eRASS J104714.1$-$435512 & 18.4~~~ & 3.4~~~~~~ & C-rich, match with nearby UV source \\
~~~~~~~~$\dots$ & J105407.37$-$650339.6 & 12.72 & [12.59:12.86] & 1eRASS J105405.9$-$650322 & 19.2~~~ & 6.7~~~~~~ & Embedded in diffuse X-ray emission \\
~~~~~~~~$\dots$ & J115211.71$-$621500.0 & 12.86 & [12.69:13.06] & 1eRASS J115210.4$-$621458 & 8.8~~~ & 3.6~~~~~~ & Embedded in diffuse X-ray emission \\
12012$-$5300 & V1110\,Cen   &  9.88 &  [9.10:10.20] & 1eRASS J120349.8$-$531642 & 19.1~~~ & 3.9~~~~~~ & Mira variable, match with nearby West source \\ 
~~~~~~~~$\dots$ & J130640.99$-$664344.3 & 15.45 & $\dots$ & 1eRASS J130639.2$-$664329 & 18.0~~~ & 4.0~~~~~~ & No obvious counterpart of X-ray source \\
15030$-$4116 & GI\,Lup      &  8.68 &  [7.10:11.36] & 1eRASS J150617.4$-$412826 & 18.3~~~ & 3.3~~~~~~ &  S Star, most likely match with Southeast \\
\multicolumn{7}{c}{} & blue source \\
~~~~~~~~$\dots$ & J152354.75$-$593054.1 & 11.99 & [11.62:12.54] & 1eRASS J152354.6$-$593104 & 10.8~~~ & 1.8~~~~~~ & Crowded field, unlikely X-IR match \\
16283$-$3447 & ~~$\dots$    & 11.78 &   $\dots$     & 1eRASS J163140.8$-$345412 & 15.2~~~ & 4.3~~~~~~ & O-rich, match nearby Southwest source \\
16387$-$2700 & AX\,Sco & 6.21 & [5.98:6.45] & 1eRASS J164149.6$-$270612 & 6.8~~~ & 5.4~~~~~~ & Optical loading, LP variable \\
~~~~~~~~$\dots$ & J165843.24$-$262141.6 & 13.00 & [12.88:13.05] & 1eRASS J165842.4$-$262149 & 13.6~~~ & 6.7~~~~~~ & No obvious counterpart of X-ray source \\
17025$-$3619 & ~~$\dots$    & 13.72 & [11.79:14.14] & 1eRASS J170553.8$-$362330 & 13.1~~~ & 5.4~~~~~~ & Extended emission, near bright source \\
~~~~~~~~$\dots$ & J171129.99$-$393304.6 & 14.08 & [12.51:15.22] & 1eRASS J171131.4$-$393254 & 19.6~~~ & 2.6~~~~~~ & Embedded in diffuse X-ray emission \\
17121$-$5200 & WRAY\,18-306 &  9.73 &   $\dots$     & 1eRASS J171601.6$-$520335 &  6.0~~~ & 5.8~~~~~~ & Extended emission, near bright source \\
~~~~~~~~$\dots$ & J171455.58$-$380600.5 & 18.36 & $\dots$ & 1eRASS J171456.3$-$380553 & 12.0~~~ & 4.2~~~~~~ & X-ray source match with HD\,155704, a G2V \\
\multicolumn{7}{c}{} & star \\
17239$-$2812 & ~~$\dots$    & 17.93 & [16.13:20.36] & 1eRASS J172705.1$-$281518 & 11.0~~~ & 3.1~~~~~~ & Extended emission, source in crowded field \\
17265$-$3230 & ~~$\dots$    & 17.88 & [16.57:18.88] & 1eRASS J172945.5$-$323237 & 12.7~~~ & 4.4~~~~~~ & O-rich, X-ray source matches Southwest star \\
\multicolumn{7}{c}{} & HD\,158270 \\ 
17364$-$4537 & V833\,Ara    & 12.41 & [11.61:15.02] & 1eRASS J174009.0$-$453905 & 12.7~~~ & 4.4~~~~~~ &  Mira variable, crowded field, no match with \\
\multicolumn{7}{c}{} & Northeast X-ray source \\
~~~~~~~~$\dots$ & J174527.09$-$343353.2 & 15.36 & [14.90:17.33] & 1eRASS J174527.8$-$343337 & 18.6~~~ & 4.1~~~~~~ & Crowded field, unlikely X-IR match \\
~~~~~~~~$\dots$ & J174829.64$-$330729.1 & 17.04 & $\dots$ & 1eRASS J174829.3$-$330715 & 13.9~~~ & 3.5~~~~~~ & Crowded field, unlikely X-IR match \\
~~~~~~~~$\dots$ & J175242.99$-$313011.4 & 15.87 & [15.02:16.29] & 1eRASS J175242.5$-$313024 & 13.8~~~ & 21.7~~~~~~ & Extended, 32 arcsec in size X-ray source \\
17534-3901   & V684\,Sco    & 12.17 & [10.78:12.82] & 1eRASS J175658.4$-$390111 & 14.9~~~ & 4.0~~~~~~ &  
Mira variable, X-ray matches Northeast optical \\
\multicolumn{7}{c}{} & source \\
\hline
\end{tabular}
\tablebib{
(1)~\citet{Schmitt2024}; 
(2)~\citet{OG2021}; 
(3)~\citet{Yu+2022}.
}
\end{table*}

\end{appendix}

\end{document}